%% file: MLPrivacy_v6.tex
\documentclass[draftcls,onecolumn,12pt]{IEEEtran}
\input{preamble.tex}
\newcommand{\tX}{{\widetilde{X}}}
\newcommand{\tcalX}{{\widetilde{\calX}}}
\newcommand{\tx}{{\widetilde{\bx}}}

\title{Data-driven Regularized Inference Privacy}
\author{Chong~Xiao~Wang and Wee~Peng~Tay,~\IEEEmembership{Senior Member,~IEEE} 
\thanks{This research is supported by the Singapore Ministry of Education Academic Research Fund Tier 2 grant MOE2018-T2-2-019 and A*STAR under its RIE2020 Advanced Manufacturing and Engineering (AME) Industry Alignment Fund – Pre Positioning (IAF-PP) (Grant No. A19D6a0053). The computational work for this article was partially performed on resources of the National Supercomputing Centre, Singapore (https://www.nscc.sg).}
\thanks{The authors are with the School of Electrical and Electronic Engineering, Nanyang Technological University, Singapore. E-mails: \texttt{{wangcx, wptay}@ntu.edu.sg}}
}

\begin{document}
\maketitle

%---------------------------------------------------------------------------%
%                           abstract and key words                          %
%---------------------------------------------------------------------------%
\begin{abstract}
Data is used widely by service providers as input to inference systems to perform decision making for authorized tasks. The raw data however allows a service provider to infer other sensitive information it has not been authorized for. We propose a data-driven inference privacy preserving framework to sanitize data so as to prevent leakage of sensitive information that is present in the raw data, while ensuring that the sanitized data is still compatible with the service provider's legacy inference system. We develop an inference privacy framework based on the variational method and include maximum mean discrepancy and domain adaption as techniques to regularize the domain of the sanitized data to ensure its legacy compatibility. However, the variational method leads to weak privacy in cases where the underlying data distribution is hard to approximate. It may also face difficulties when handling continuous private variables. To overcome this, we propose an alternative formulation of the privacy metric using maximal correlation and we present empirical methods to estimate it. Finally, we develop a deep learning model as an example of the proposed inference privacy framework. Numerical experiments verify the feasibility of our approach.
\end{abstract}


%-------------------------------Section----------------------------------%
\section{Introduction} 
\label{sect:intro}
In recent years, artificial intelligence and machine learning have found significant commercial applications for complex tasks such as object recognition, speech recognition and language translation. Big tech companies such as Google, Microsoft and Amazon have provided easy access to efficient and sophisticated information processing supported by large data storage, high computing power and intelligent software. While this technology benefits humanity, it also carries privacy leakage risks, which is a major concern that is gaining attention. When a user sends data to a service provider for it to perform inference for an authorized task, potential adversaries (including the service provider) may exploit the raw data to glean sensitive information. Studies \cite{FirGolElo:J14,AbaNinHer:J16} have shown that users' personal information such as sexual orientation or political affiliation can be accurately inferred from their activities on social networking platforms. The situation is further aggravated by data breach incidents, e.g., $1.5$ million members of Singapore's largest healthcare group had their personal data compromised in $2018$ \cite{WiKi:2018}. Due to privacy concerns, data owners may feel reluctant to share their data with others in collaborative learning where data from various data owners are collected and analyzed. E.g., social network users in a community may share their personal opinions or experiences about different products over time. While the aggregated data can be useful feedback for a company to improve its own products, with a database recording the preferences of each user for multiple products over a period of time, one can infer personal traits and other sensitive attributes like the gender and income level of a user. Therefore, data sanitization is imperative in the big data era for protecting privacy in data sharing and cloud computing. 

Privacy can be categorized into data privacy and inference privacy. Data privacy refers to protecting the access to the raw data while inference privacy refers to the prevention of inferring sensitive information \cite{CalFaw:C12,SunTay:J19b}. We are concerned with preserving \emph{inference privacy} in this paper. Differential privacy\cite{NyPap:J14} is widely used as a privacy metric both in data privacy and inference privacy, which ensures the indistinguishability of the query records in a database. The other privacy metrics extensively used in protecting inference privacy include mutual information (entropy) privacy, average information leakage and maximum information leakage \cite{CalFaw:C12}. The notion of information privacy was proposed by \cite{CalFaw:C12} and explored by \cite{SunTay:J19a}. The relationships among these metrics are studied in \cite{WanYinZha:J16,SunTay:J19b}. In terms of data sanitization mechanism, two commonly used methods are: (1) compressive privacy \cite{Kung:J2017,TseBoWu:J2020,SonWanTay:J19} that transforms raw measurements to a lower dimensional space, and (2) probabilistic mapping \cite{CalFaw:C12,SanRajPoo:J2013,LiaSanTan:J2017} that adds randomness to perturb raw measurements. We adopt the latter as the data sanitization scheme in this paper.

\subsection{Related Work}
Considering the problem of disclosing a random variable $X$ that is correlated with a private variable $S$, the authors of \cite{CalFaw:C12,MakSalFaw:C14} proposed to release a sanitized version $Z$ of $X$ via probabilistic mapping, by minimizing the mutual information between $Z$ and $S$ given a maximum allowable distortion between $Z$ and $X$. Following the same setup, the papers \cite{WanCal:C2017,WanVoCal:J2019} investigated the utility-privacy tradeoff from an estimation-theoretic viewpoint based on the techniques developed in \cite{CalMak:J2017}. The papers \cite{AsoAlaLin:C16,AsoDiaLin:J17} formulated the utility-privacy tradeoff in terms of the minimum mean-squared error (MMSE) and \cite{SanRajPoo:J2013} presented an information-theoretic framework that provides tight bounds for the utility and privacy metrics. Moreover, the papers \cite{CalMak:C2015,RasGun:J2017} quantified the utility achievable (mutual information between $Z$ and $X$) under perfect privacy (i.e., $Z$ and $S$ are independent). In a different setting, \cite{LiaSanTan:J2017} maximizes the relative entropy between pairs of distributions of the sanitized data for hypothesis testing while requiring that the mutual information between the sanitized data and raw data is small. Recently, \cite{WanSonTay:J2020,SonWanTay:J19,SonWanTay:C18,WanSon:C2018} studied the problem of preserving the estimation privacy of a set of private parameters in a linear system. Linear compression and noise perturbation mechanisms are used as sanitization techniques. However, applying their proposed frameworks in practice remains challenging as the underlying assumption of known data distribution or system model is often untenable. Furthermore, for all the aforementioned works, to implement their proposed privacy-preserving frameworks require changing the service provider's backend analytics systems to be compatible with the sanitized data, which is very different from the raw data.

Several works have explored data-driven approaches from different perspectives. The paper \cite{Kung:J2017} introduced a dimension reduction subspace approach. As an extension, the paper \cite{ChaChaKun:C17} studied the design of privacy-preserving mechanisms based on compressive privacy and multi-kernel methods. The reference \cite{SunTayHe:J18} proposed a nonparametric learning approach to design privacy mappings to distort sensors' observations. In a similar setting, the paper \cite{HeWee:J2019} proposed a multilayer nonlinear processing procedure to distort sensors' data. However, these works do not cater for a wide range of privacy problems since \cite{Kung:J2017} focuses on linear utility and privacy spaces and \cite{ChaChaKun:C17,SunTayHe:J18,HeWee:J2019} target at classification problems only. More closely related works are \cite{ChrKevYuj:J2015,DanShuRob:J2018,RicYuKev:C2013}, which use auto-encoders to learn representations that are invariant to certain sensitive factors in the data while retaining as much of the remaining information as possible. However, these works imposed certain data model assumptions and cannot be generalized to broader privacy models. 

Recently, the papers \cite{ChoPetXia:J2018,ChoPetXia:J2017,TriWanIsh:C19} presented a data-driven framework called generative adversarial privacy (GAP) by formulating the privacy mechanism as a constrained minimax game between a privatizer and an adversary. The utility measure is based on the pixel distortion between the raw data and sanitized data. This is heuristic and restrictive as larger distortion does not imply less utility. The method also distorts the sanitized data significantly so that legacy systems built for the raw data does not perform well or cannot be applied to the sanitized data directly. Similarly, the paper \cite{TseBoWu:J2020} proposed a compressive privacy generative adversarial framework for generating compressing representations that retain utility comparable to the state-of-the-art when defending against reconstruction attacks. Even though it takes into account multiple adversarial strategies for more robust evaluation of adversarial reconstruction attacks, it does not give an assurance that it can withstand other adversarial attacks.

All the aforementioned works perform data sanitization by mapping the raw data into a different space, which prevents legacy inference systems from working with the sanitized data. This paper proposes the use of regularization techniques to minimize the disparity between the distributions of the sanitized data and raw data, making our privacy preserving framework compatible with a service provider's legacy learning systems. Our proposed approach thus incurs no additional cost for the service provider. 

\subsection{Our Contributions}
We propose a Data-driven Regularized Inference Privacy (DRIP) framework to prevent the leakage of private information when sharing data. DRIP ensures that the sanitized data is compatible with the service provider's legacy learning systems. 

A privacy function measures the dependence of the private variable on the sanitized data and should achieve its optimum when the sanitized data and private variable are independent. From this point of view, an independence criterion is a reasonable privacy metric choice. The utility of the sanitized data can be regarded to be high if the output response from the service provider is similar to the response when fed with the raw data. When a specific public variable is predefined by the user, utility can be related to the loss of an appropriately chosen inference model on that public variable. Otherwise, instead of choosing an average distance function between the raw and sanitized data, we formulate the utility from an information theoretic viewpoint. Our main contributions are the following:
\begin{enumerate}
\item We approximate the mutual information between the raw data and sanitized data as a measure of utility and incorporate the notion of average information leakage as the privacy metric in the DRIP framework based on a variational method. This framework is called DRIP-Var.
\item We introduce domain regularization techniques to ensure that the sanitized data is compatible with a legacy inference system by forcing the marginal distributions between the sanitized data and the raw data to be similar. 
\item To avoid uncertainty in errors introduced by the variational method in cases where the data distribution is difficult to approximate and to provide a way to handle continuous private variables efficiently, we propose an alternative privacy function using maximal correlation. We develop several methods to estimate maximal correlation. This framework is called DRIP-Max.
\item We illustrate our proposed frameworks using a deep learning architecture and verify its performance on an image dataset \cite{ZhaSonQi:C2017} and a diabetes dataset \cite{ErfHasJoh:J2004}. 
\end{enumerate}

The rest of this paper is organized as follows. In \cref{sect:problem_statement}, we present our problem formulation and assumptions. In \cref{sect:var_method}, we discuss the privacy and utility metrics for DRIP-Var as well as the legacy compatibility of our approach. In \cref{sect:max_corr}, we present the maximal correlation approach in DRIP-Max as an alternative privacy metric and discuss methods to estimate it. In \cref{sect:model_dp}, we illustrate our DRIP framework using a deep learning architecture. In \cref{sect:experiment}, we evaluate the proposed framework on real-world datasets. \cref{sect:conclusion} concludes the paper.

\emph{Notations:}
We use capital letters like $X$ to denote random variables or vectors, lowercase letters like $x$ for deterministic scalars, and boldface lowercase letters like $\bx$ for deterministic vectors. Throughout this paper, we assume that every random variable has a (generalized) probability density function (pdf) (for discrete random variables, this specializes to a probability mass function). We use $p_X(\cdot)$ to denote the pdf of $X$, and $p_{X|Y}(\cdot|\cdot)$ to denote the conditional pdf of $X$ given $Y$. We use $X\sim p_X$ to denote that the random variable $X$ follows a pdf $p_X$. $\E$ is the expectation operator and we use $\E_X$ to emphasize that the expectation is \gls{wrt} $X$. We denote $H(X)$ as the entropy of the random variable $X$, and $I(X;Y)$ as the mutual information between $X$ and $Y$. We denote $D(\cdot,\cdot)$ as the Kullback–Leibler divergence. The Gaussian distribution with mean $\mu$ and variance $\sigma^2$ is denoted as $\N{\mu}{\sigma^2}$.

%-------------------------------Section----------------------------------%
\section{Problem Formulation}
\label{sect:problem_statement}
\begin{figure}[!htb]
	\centering
	\includegraphics[scale=1.1]{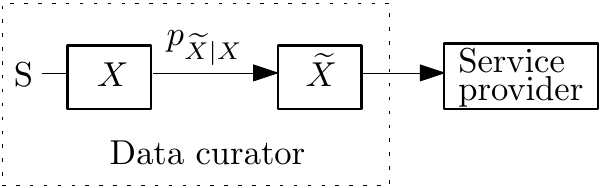}
	\caption{Privacy framework.}
	\label{fig:privacy_diagram}
\end{figure}

We consider a random variable $X\in\calX$ on which a private variable $S\in\calS$ is statistically dependent. A data curator obtains an instance of $X$ and intends to send it to a service provider (e.g., a cloud service) who can help her to infer some public information from $X$. However, she is also concerned that the service provider can exploit the raw data $X$ to infer the private variable $S$. For example, users may wish not to reveal their ethnicity or gender when uploading their facial images to a server for verification purposes. To protect $S$, the data curator sanitizes the raw data $X$ to reduce the amount of private information, while maximally retaining the rest of the information to preserve utility. As sketched in \cref{fig:privacy_diagram}, the data curator sends a sanitized version of data $\tX\in\tcalX$ to the service provider.

Given a training dataset $\calD=\{(\bx_i,\bs_i)\}_{i=1}^N$, where $(\bx_i,\bs_i)$ is an instance of the random vector $(X,S)$ with joint probability density $p_{X,S}$, our task is to learn a privacy-preserving probabilistic function that maps the raw data $X$ to sanitized data $\tX$:
\begin{align}\label{eq:sanitizer}
p_{\tX|X}: \calX\times\tcalX\mapsto[0,1].
\end{align}
In theory, the search space of the probabilistic mapping $p_{\tX|X}$ is not constrained, which makes finding a mapping that optimizes the tradeoff between utility and privacy computationally expensive in some cases. To make the problem more tractable, we restrict to a rich class of mappings by relating the mappings to a parameterized differentiable function via \emph{reparameterization}: Under certain mild conditions (cf. Section 2.4 in \cite{KinWel:J2013}) for a chosen posterior $p_{\tX|X}$, we can reparameterize the random variable $\tX$ using a differentiable transformation of an auxiliary continuous random variable $\Xi$ independent of $X$:
\begin{align} \label{eq:reparam}
\tX=f^{\btheta}(X,\Xi),\ \Xi\sim p_\Xi,
\end{align}
where $f^{\btheta}$ is a function parameterized by $\btheta$ and differentiable \gls{wrt} $\btheta$. 
%The inverse probability integral transform can serve as an intuition behind the reparameterization, which states: if the random variable $\Xi$ has a uniform distribution on $[0,1]$ and if the random variable $\tX$ has a cumulative distribution $F(\tX)$, then the random variable $F^{-1}(\Xi)$ has the same distribution as $\tX$. 
In practice, we let $f^{\btheta}$ be represented by a neural network because of its ability to approximate a rich class of functions efficiently. The reparameterization allows us to apply the stochastic gradient-descent method to optimize over the sanitization parameter $\btheta$. For the sake of notational clarity, we rewrite $p_{\tX|X}$ as $p_{\tX|X}^{\btheta}$.

We seek to optimize over $\btheta$ to find an optimal sanitizer to achieve a desirable utility and privacy tradeoff:
\begin{align} \label{opt:obj}
\max_{\btheta}\scL(\btheta)-\lambda_1\scP(\btheta)-\lambda_2\scR(\btheta),
\end{align}
where $\lambda_1$ and $\lambda_2$ are positive constants, $\scL(\btheta)$ is the utility function representing the public information retained in the sanitized data $\tX$, $\scP(\btheta)$ is the privacy function quantifying the amount of statistical information of $S$ remaining in $\tX$, and $\scR(\btheta)$ denotes a regularization term. As the inference system of a service provider (which we call a legacy system) is designed to work with the raw unsanitized data $X$, designing the privacy-preserving function $p_{\tX\mid X}$ so that the sanitized data $\tX$ is compatible with the legacy system would be of practical importance. This allows the service provider to save costs and time as it can reuse the legacy system for the sanitized data. Moreover, it also need not maintain two different systems, one for users with no privacy sanitization and one for users with sanitization. The regularization term $\scR(\btheta)$ is designed to ensure that the marginal distributions of the sanitized data and raw data ($p_{\tX}$ and $p_X$) are close to each other.

\begin{figure}[!htb]
	\centering
	\includegraphics[scale=1.1]{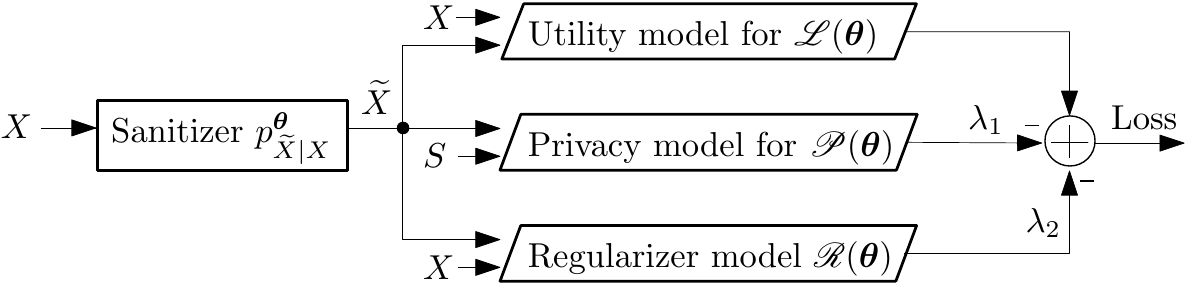}
	\caption{Machine learning DRIP architecture.}
	\label{fig:put_diagram}
\end{figure}

We represent the sanitizer, privacy and utility functions, and regularizer in our DRIP framework using machine learning models, each with their own learnable parameters. This allows us to back-propagate the gradient
\begin{align}\label{eq:gradient}
\left(
\frac{\partial\scL(\btheta)}{\partial\tX} 
-\lambda_1\frac{\partial\scP(\btheta)}{\partial\tX}
-\lambda_2\frac{\partial\scR(\btheta)}{\partial\tX}
\right)\frac{\partial\tX}{\partial\btheta}
\end{align}
through the sanitizer network (see \cref{fig:put_diagram}) to update the parameter $\btheta$ so as to learn the optimal sanitizer.
%\blue{Note $\tX$ is a function of the random variables $X$ and $\Xi$. Therefore, $\frac{\partial\tX}{\partial\btheta}$ is always evaluated at a fixed $X$ and $\Xi$ in order to obtain a numerical value.}

\begin{Remark}
In a decentralized setting as illustrated in \cref{fig:decentralized_privacy_diagram} where a fusion center collects data from multiple data curators, each data curator $i$ sanitizes its raw data $X_i\in\calX_i$ to obtain a sanitized version $\tX_i\in\tcalX_i$ and sends $\tX_i$ to the fusion center to prevent the fusion center from inferring the private variable $S$, while allowing it to infer some public information (utility). We assume that there is no message exchange between data curators during the sanitization process as this is prone to privacy attacks. As a consequence, we let each data curator $i$ adopt an independent local data sanitization $p_{\tX_i|X_i}: \calX_i\times\tcalX_i\mapsto[0,1]$, which results in a decentralized sanitization scheme:
\begin{align*}
p_{\tX|X}=\prod_{i=1}^N p_{\tX_i|X_i} : \calX\times\tcalX\mapsto[0,1],
\end{align*} 
where $\tX=\left[\tX_1,\ldots,\tX_N\right]$ and $X=\left[X_1,\ldots,X_N\right]$. We reparameterize $\tX_i$ as $\tX_i=f_i(X_i,\Xi_i)$ where $\Xi_i\sim p_{\Xi_i}$ is a random variable independent of $X_i$ for all $i=1,\ldots,N$ as well as $\Xi_j$ for $j\neq i$. This configuration can be viewed as a special case of our framework.
\begin{figure}[!htb] 
	\centering 
	\includegraphics[scale=1.2]{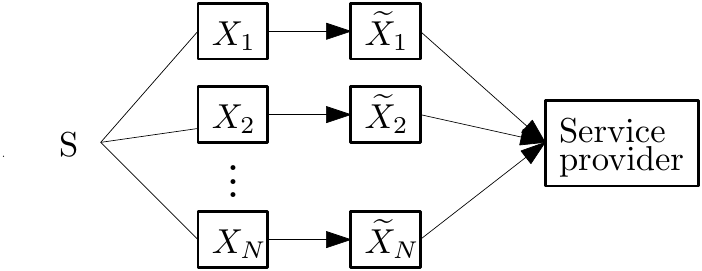}
	\caption{Decentralized DRIP framework.}
	\label{fig:decentralized_privacy_diagram}
\end{figure}
\end{Remark}
\begin{Remark}\label{rem:public_variable}
Hitherto our problem formulation does not include a public variable. There are cases where users are interested in inferring a specific public variable only (denoted as $U$) while preserving the privacy of the sensitive variable $S$. In such cases, we simply let the utility function be the negative minimized loss of a state-of-the-art inference model for $U$. For example, if $U$ can be inferred from $X$ by using a regression model, we then let
\begin{align*}
\scL(\btheta)=-\min_{h}\E[\left(h(\tX)-U\right)^2].
\end{align*}
In this manner, as demonstrated in our experiments in \cref{sect:results_max_corr}, we can achieve utility of the sanitized data comparable to that of the unsanitized data. In another case that is the converse of our problem statement, users may be interested to infer a specific public variable $U$, without specifying the private variable $S$. This can be realized by replacing $S$ with $U$ and reversing the sign of the utility and privacy functions in \cref{opt:obj}.
\end{Remark}
%

%-------------------------------Section----------------------------------%
\section{Variational Method for Utility and Privacy Metrics}
\label{sect:var_method}

In this section, we formulate the utility and privacy metrics using the variational method \cite{KinWel:J2013}. We discuss how to regularize the domain of the sanitized data to make it compatible with a service provider's legacy system.

\subsection{Utility Function}
When sanitizing the raw data $X$ to remove the private information $S$, we seek to retain as much of the non-sensitive information as possible to maintain the fidelity of the sanitized data so that it can still be useful for the public tasks. A widely-used utility metric in the privacy literature is mutual information, which measures the statistical dependence between two random variables ($X$ and $\tX$ in our case). While the expression of mutual information is concise and elegant in theory, its computation is challenging as it requires the data distribution to be known, which is often not possible in practice. To overcome this, we propose to use the variational method \cite{KinWel:J2013} to approximate the mutual information between $X$ and $\tX$ and use it as the utility function. 

Consider the privacy-preserving probabilistic mapping $p_{\tX|X}^{\btheta}$ with a fixed parameter $\btheta$. 
The mutual information between $X$ and $\tX$ is
\begin{align*}
I(X;\tX)=H(X)-H(X|\tX).
\end{align*}
Since the raw data distribution $p_X$ is implicitly specified, $H(X)$ is a constant. Therefore, we are only interested in estimating
\begin{align*}
H(X|\tX)=-\E_{(X,\tX)\sim p_{X,\tX}^{\btheta}}[\log p_{X|\tX}(X|\tX)],
\end{align*}
where $p_{X,\tX}^{\btheta}(\bx,\tx)=p_{\tX|X}^{\btheta}(\tx|\bx)p_X(\bx)$. Here $p_{\tX|X}^{\btheta}$ and $p_{X|\tX}$ bear some resemblance to the notion of the encoder and decoder of an autoencoder, respectively. While an autoencoder encodes $X$ to a latent representation, we wish to render $\tX$ as an approximation of $X$ with private information removed. To avoid confusion, we refer to $p_{X|\tX}^{\btheta}$ as the posterior of the raw data $X$. Note that computing $p_{X|\tX}$ is intractable since $p_X$ is unknown to us. Even if $p_X$ is given, it is not practical to derive $p_{X|\tX}$ via Bayes’ theorem because the computation of the evidence $p_{\tX}$ involves an integral over the sample space of $X$, which can be a high dimensional space. To overcome this difficulty, we employ the variational method to obtain an upper bound for $H(X|\tX)$.
 
Let $q_{X|\tX}^{\bphi}$ be a proposed variational distribution parametrized by $\bphi$, representing a set of candidate functions used for approximating the true distribution $p_{X|\tX}$. We have
\begin{align*}
-H(X|\tX)
&= \E{(X,\tX)\sim p_{X,\tX}^{\btheta}}[\log p_{X|\tX}(X|\tX)] \\
%&=\E{(X,\tX)\sim p_{X,\tX}^{\btheta}}[\log q_{X|\tX}^{\bphi}(X|\tX)] + \E{(X,\tX)\sim p_{X,\tX}^{\btheta}}[\log \frac{p_{X|\tX}(X|\tX)}{q_{X|\tX}^{\bphi}(X|\tX)}] \\
&=\E{(X,\tX)\sim p_{X,\tX}^{\btheta}}[\log q^{\bphi}_{X|\tX}(X|\tX)]
	+\E_{\tX\sim p_{\tX}}[D\left(p_{X|\tX}(\cdot|\tX)||q^{\bphi}_{X|\tX}(\cdot|\tX)\right)],
\end{align*}
where $D(\cdot,\cdot)$ denotes the Kullback–Leibler divergence. Because of the non-negativity of the Kullback–Leibler divergence, we have
\begin{align*}
\E_{\tX\sim p_{\tX}}[D\left(p_{X|\tX}(\cdot|\tX)||q^{\bphi}_{X|\tX}(\cdot|\tX)\right)]\geq 0.
\end{align*}
Therefore, $\E{(X,\tX)\sim p_{X,\tX}^{\btheta}}[\log q^{\bphi}_{X|\tX}(X|\tX)]$ forms a lower bound for $-H(X|\tX)$. The equality holds if and only if $q_{X|\tX}^{\bphi}(X|\tX)=p_{X|\tX}(X|\tX)$ almost surely. We maximize this lower bound over the variational parameter $\bphi$ to approximate $-H(X|\tX)$. Recall that $\tX$ is determined by the sanitization parameter $\btheta$. Let the utility function be 
\begin{align}\label{utility}
\scL(\btheta)=\max_{\bphi}\scL(\bphi;\btheta),
\end{align} 
where
\begin{align*}
\scL(\bphi;\btheta)=\E{(X,\tX)\sim p^{\btheta}_{X,\tX}}[\log q^{\bphi}_{X|\tX}(X|\tX)].
\end{align*}
The \emph{reparameterization} in \cref{eq:reparam} is useful for our formulation because it can be used to rewrite the expectation \gls{wrt} $p^{\btheta}_{X,\tX}$ as an expectation \gls{wrt} $p_X,p_\Xi$. Then, $\scL(\bphi;\btheta)$ can be rewritten as
\begin{align}\label{eq:va_utility}
\scL(\bphi;\btheta)=\E_{X\sim p_X,\Xi\sim p_\Xi}[\log q^{\bphi}_{X|\tX}(X|\tX)],
\end{align}
where $\tX=f^{\btheta}(X,\Xi)$. We can use Monte Carlo methods to estimate \cref{eq:va_utility} by drawing samples from $p_X$ and $p_\Xi$ independently. Samples of $p_X$ are given as training data and $p_\Xi$ is often chosen to be a common probability distribution such as the Gaussian or uniform distributions.

\subsection{Privacy Function} \label{sect:var_privacy}

To prevent an adversary from inferring the private variable $S$ from the sanitized data $\tX$, we adopt the notion of \emph{average information leakage} \cite{CalFaw:C12} to form the privacy function
\begin{align}\label{ItXS}
\E_{\tX\sim p_{\tX}}[D\left(p_{S|\tX}(\cdot|\tX),p_S(\cdot)\right)]
=H(S)-H(S|\tX),
\end{align}
which attains the minimum when $S$ and $\tX$ are independent. Because \cref{ItXS} can not be computed analytically without knowledge of the joint distribution $p_{S,\tX}$, we use the variational method \cite{KinWel:J2013} to obtain an approximation. Let a variational distribution $q^{\btau}_{S|\tX}$ parameterized by $\btau$ act as a surrogate for the true distribution $p_{S|\tX}$. We have
\begin{align}
-H(S|\tX)
&=\E_{(S,\tX)\sim p_{S,\tX}}[\log p_{S|\tX}(S|\tX)] \nn 
&=\E_{(S,\tX)\sim p_{S,\tX}}[\log q^{\btau}_{S|\tX}(S|\tX)]
+ \E_{\tX\sim p_{\tX}}[\KLD{p_{S|\tX}(\cdot|\tX)}{q^{\btau}_{S|\tX}(\cdot|\tX)}]\nn
&\geq \E_{(S,\tX)\sim p_{S,\tX}}[\log q^{\btau}_{S|\tX}(S|\tX)], \label{ineq:HStX}
\end{align}
where equality holds if and only if $q^{\btau}_{S|\tX}=p_{S|\tX}$ almost surely. Because $\tX - X - S$ forms a Markov chain, we further have
\begin{align}
&\E_{(S,\tX)\sim p_{S,\tX}}[\log q^{\btau}_{S|\tX}(S|\tX)] \nn
&=\E_{(S,X,\tX)\sim p_{S,X,\tX}}[\log q^{\btau}_{S|\tX}(S|\tX)] \nn
&=-\E_{(X,\tX) \sim p^{\btheta}_{X,\tX}}[H\left(p_{S|X}(\cdot|X),q^{\btau}_{S|\tX}(\cdot|\tX)\right)],\label{va1}
\end{align}
where $H(p,q)=-\E_{Z\sim p}[\log q(Z)]$ is the cross-entropy function.
From \cref{ItXS,ineq:HStX}, we obtain
\begin{align}\label{ineq:ItXS}
I(S;\tX) \geq H(S) + \scP(\btau;\btheta),
\end{align}
where
\begin{align} \label{eq:va_privacy}
\scP(\btau;\btheta)
=-\E_{X\sim p_X,\Xi\sim p_{\Xi}}[H\left(p_{S|X}(\cdot|X),q^{\btau}_{S|\tX}(\cdot|\tX)\right)]
\end{align}
is \cref{va1} after applying the reparametrization trick and $\tX=f^{\btheta}(X,\Xi)$. Using the lower bound \cref{ineq:ItXS} and ignoring the constant $H(S)$, we obtain an approximate privacy function as
\begin{align}\label{va_approx}
\scP(\btheta)=\max_{\btau}\scP(\btau;\btheta).
\end{align} 
To make use of \cref{eq:va_privacy}, we need to assume that $p_{S|X}$ can be obtained from the training data in the form of soft or hard labels. An interesting observation is that the adversarial training for classification problems using softmax activation in \cite{ChoPetXia:J2018} can be considered as a special case of this procedure (cf.\ \cref{sect:model_dp} for an illustration).

Notice that we are using a lower-bound to approximate the mutual information between the private variable and the sanitized data for the privacy function. The real average information leakage of the private variable given by $I(S;\tX)$ is always more serious than its lower bound in \cref{ineq:ItXS}, with the discrepancy quantified as
\begin{align*}
H(S|\tX)-\scP(\btheta)
=\min_{\btau}\E_{\tX\sim p_{\tX}}[D(q_{S|\tX}^{\btau}(\cdot|\tX)||p_{S|\tX}(\cdot|\tX)] \geq 0.
\end{align*}
Only when the proposed variational distribution $q_{S|\tX}^{\btau}$ is able to approximate the true distribution $p_{S|\tX}$ well, we are able to maintain a privacy level within a small margin of error. As a consequence, in cases where a close approximation of the true distribution cannot found due to a lack of prior knowledge or domain expertise, the variational method may lead to unacceptable privacy leakage. In contrast, we are not concerned with the utility function where a lower-bound approximation is also used as we always get better utility in reality than the approximation given by the variational method. In addition, to make the analytical computation of \cref{eq:va_privacy} tractable, we require $S$ to have finite support. These considerations prompt us to seek for an alternative privacy metric that is more robust and can tackle continuous private variables, i.e., the maximal correlation in \cref{sect:max_corr}.

\subsection{Legacy Compatibility}
Since mutual information is insensitive to the symbols used to represent the random variables, $\tX$ produced by maximizing \cref{utility} subject to a constraint on \cref{va_approx} can be very different from $X$. In many applications, we desire the sanitized data $\tX$ to be similar to $X$. For example, when sanitizing face images to remove private information such as gender and ethnicity, we expect the sanitized data to still be face images. This can not be guaranteed by mutual information. To achieve this, we force the marginal distribution of $\tX$ to be close to the marginal distribution of $X$. A heuristic method to achieve this is to minimize the Euclidean distance $\E\lVert{X-\tX}\rVert$, which however shows poor performance for unstructured data such as images in our experiments. In this paper, we propose two techniques to make $p_{\tX}$ similar to $p_X$: (1) bounding the maximum mean discrepancy\cite{GreBor:J2006}, which is a metric on the space of probability measures, and (2) domain adaption\cite{ZhuPar:J2018,MurKolKri:C2018} based on deep learning, which learns the mapping between a source space and a target space using a training set. The former is efficacious in coping with data of simple structure while the latter is favored when processing high-dimensional data.

\subsubsection{Maximum mean discrepancy (MMD)} The MMD measures the disparity between distributions \cite{Mul:J1997}:
\begin{align*}
\mathrm{MMD}(\calF,p_X,p_{\tX})=\sup_{f\in\calF} \braces*{\E_{X\sim p_X} f(X)-\E_{\tX\sim p_{\tX}} f(\tX)},
\end{align*}
where $\calF$ is a class of witness functions $f:\calX\mapsto\bbR$ that determines the probability metric. In particular, the paper \cite{SriFukLan:J2011} sets the witness function class $\calF$ to be the unit ball in a reproducing kernel Hilbert space (RKHS). Let $\calH$ be a RKHS with kernel $K(\cdot,\cdot)$ and $\Phi(\bx)=K(\cdot,\bx)$ be the feature mapping that maps $\bx\in\calX$ to elements in $\calH$. Denote $\norm{\cdot}_\calH$ as the norm induced by the kernel inner product $\ip{\cdot}{\cdot}_\calH$. A RKHS has the reproducing property $f(\bx)=\langle{\Phi(\bx),f}\rangle_\calH$ for all $f\in\calH$ \cite{ManAmb:J2015}. This leads to an analytical expression of squared MMD under RKHSs \cite{GreBorRas:J2012}:
\begin{align}
&\mathrm{MMD}^2(\calH,p_X,p_{\tX}) \nn
&=\sup_{\norm{f}_\calH \leq 1} \braces*{\E_{X\sim p_X} f(X)-\E_{\tX\sim p_{\tX}} f(\tX)}^2 \nn
&=\sup_{\norm{f}_\calH \leq 1} \braces*{\ip{\E_{X\sim p_X}\Phi(X)-\E_{\tX\sim p_{\tX}} \Phi(\tX)}{f}_\calH}^2 \nn
&=\norm{\E_{X\sim p_X}\Phi(X)-\E_{\tX\sim p_{\tX}}\Phi(\tX)}_\calH^2\nn
&=\E_{X,X'\sim p_X}[K(X,X')] - 2 \E_{X\sim p_X,\tX\sim p_{\tX'}}[K(X,\tX)] + \E_{\tX,\tX'\sim p_{\tX}}[K(\tX,\tX')], \label{eq:MMD2}
\end{align}
where we have assumed that $\E_{X\sim p_X}\Phi(X)$ and $\E_{\tX\sim p_{\tX}}\Phi(\tX)$ exist. This embedding encodes a probability measure as a mean element in a RKHS and minimizing the MMD can be viewed as matching all of the moments of $p_X$ and $p_{\tX}$. For a characteristic (universal) kernel such as the Gaussian kernel, MMD is zero if and only if $p_X=p_{\tX}$ \cite{GreBorRas:J2012}. Using the empirical form of \cref{eq:MMD2}, we then obtain an unbiased estimator of the squared MMD that distinguishes two distributions from a set of training data $\{(\bx_i,\tx_i)\}_{i=1}^N$:
\begin{align} \label{eq:rgl_mmd}
\begin{split}
\widehat{\mathrm{MMD}}_K^2(p_X,p_{\tX})
&=\frac{1}{N^2}\sum_{i=1}^N\sum_{j=1}^N K(\bx_i,\bx_j)
-\frac{2}{N^2}\sum_{i=1}^N\sum_{j=1}^N K(\tx_i,\bx_j) \\
&+\frac{1}{N^2}\sum_{i=1}^N\sum_{j=1}^N K(\tx_i,\tx_j),
\end{split}
\end{align}
which is differentiable \gls{wrt} the sanitization parameter $\btheta$ if the kernel $K$ is differentiable. The paper \cite{LiChaChe:J2017} considers a composition characteristic kernel $K\circ f=K(f(\cdot), f(\cdot))$, where $K(\cdot,\cdot)$ is a pre-specified kernel and $f$ is an injective function. To add robustness to the estimator, it optimizes over a set of possible characteristic kernels: $\max_{f}\widehat{\mathrm{MMD}}_{K\circ f}^2(p_X,p_{\tX})$.  A similar approach can be found in \cite{LiSweZem:J2015}. As the amount of data required to produce a reliable MMD estimator grows with the dimensionality of the data, using $f$ to encode data into lower spaces can be beneficial for high-dimensional data that exist on a low-dimensional manifold such as visual data.

\subsubsection{Domain adaption} Let random variables $A\in\calA$ and $B\in\calB$ be associated with probability distributions $p_A$ and $p_B$, respectively. Domain adaption aims to learn a function $G$ that transforms a data in the source domain $\calA$ to the target domain $\calB$ such that the distribution of $G(A)$ is indistinguishable from $p_B$. The objective can be written as
\begin{align*}
&\min_G\max_{D_B}\E_{B\sim p_B}[\log(D_B(B))]+\E_{A\sim p_A}[\log(1-D_B(G(A))],
\end{align*}
where $D_B : \calB\cup{G(\calA)} \mapsto [0,1]$ is an adversarial discriminator seeking to differentiate the transformed data $G(A)$ from $B$ while the function $G$ aims to make the distribution of $G(A)$ resemble that of $B$. The discriminator and transformation function are trained against each other until an equilibrium is reached. Motivated by this idea, we introduce a regularity term targeting at $\tX\sim p_X$. Denote
\begin{align} \label{eq:rgl_da}
\scR(D_X;\btheta)=\E_{X\sim p_X}[\log(D_X(X))] +\E_{\tX\sim p_{\tX}}[\log\parens*{1-D_X(\tX)}],
\end{align}
where $D_X :\calX\cup\tcalX\mapsto [0,1]$ is a discriminator yielding the likelihood that the input is sampled from $p_X$. We optimize the discriminator $D_X$ to obtain 
\begin{align}\label{va_rgl}
\scR(\btheta)=\max_{D_X}\scR(D_X;\btheta),
\end{align}
which we then minimize over $\btheta$.

\subsection{Overall DRIP-Var Framework}

In summary, by substituting \cref{utility,va_approx,va_rgl} into \cref{opt:obj}, the overall objective function can be rewritten as a max-min problem:
\begin{align}\label{opt:va_obj}
\max_{\btheta,\bphi}\min_{\btau,D_X}\scL(\bphi;\btheta)-\lambda_1\scP(\btau;\btheta)-\lambda_2\scR(D_X;\btheta).
\end{align}
The regularization term $\scR(D_X;\btheta)$ can be replaced by a maximization of \cref{eq:rgl_mmd} if the MMD is used as the regularizer. The max-min problem \cref{opt:va_obj} attempts to maximize the utility-privacy tradeoff with a regularizer guiding the sanitized data to be in a sensible domain. In order to optimize \cref{opt:va_obj} using the gradient descent method, the proposed sanitizer $p_{\tX|X}^{\btheta}$ and variational distributions $q^{\bphi}_{X|\tX}$ and $q^{\btau}_{S|\tX}$ are required to take certain forms and constraints, which are summarized as follows:
\begin{enumerate}[(i)]
	\item The sanitizer $p_{\tX|X}^{\btheta}$ in \cref{eq:sanitizer} is determined by $\tX=f^{\btheta}(X,\Xi)$, where $\Xi\sim p_\Xi$, and $f^{\btheta}(X,\Xi)$ is differentiable \gls{wrt} $\btheta$. $\tX$ has the same dimension as $X$.
	\item The function $q^{\bphi}_{X|\tX}:\calX\times\tcalX\mapsto [0,1]$ used in the utility function \cref{utility} is such that $\int_{\bx\in\calX}q^{\bphi}_{X|\tX}(\bx|\tx)=1$ for any $\tx\in\tcalX$, and $q^{\bphi}_{X|\tX}(\bx|\tx)$ is differentiable \gls{wrt} $\bphi$ and $\tx$. 
	\item The function $q^{\btau}_{S|\tX}:\calS\times\tcalX\mapsto [0,1]$ used in the privacy function \cref{va_approx}  is such that $\sum_{\bs\in\calS}q^{\btau}_{S|\tX}(\bs|\tx)=1$ for any $\tx\in\tcalX$, and $q^{\btau}_{S|\tX}(\bs|\tx)$ is differentiable \gls{wrt} $\btau$ and $\tx$.
\end{enumerate}
%

%-------------------------------Section----------------------------------%
%
\section{Maximal Correlation as Privacy Metric}
\label{sect:max_corr}
In this section, we formulate the privacy function from an estimation perspective using the concept of maximal correlation. Maximal correlation measures the statistical dependency between two random variables by using two appropriate functions that map these random variables to real-valued spaces. The estimation of maximal correlation does not explicitly depend on the underlying probability distribution, while the variational method in \cref{sect:var_privacy} needs to approximate the distribution. Moreover, maximal correlation can handle both continuous and discrete private variables, while the variational method has difficulty in dealing with continuous private variables. We develop methods to estimate the maximal correlation between random variables.

\subsection{Definition and Properties}

Maximal correlation was introduced by Hirschfeld \cite{Hir:J1935} and Gebelein \cite{Gel:J1941}, and later studied by R\'enyi \cite{Ren:J1959}. It measures the cosine of the angle between two linear subspaces of zero mean, square integrable real-valued random variables. Given two jointly distributed random variables $Y\in\calY$ and $Z\in\calZ$, the maximal correlation of $Y$ and $Z$ is defined as follows.
\begin{Definition}[The Hirschfeld-Gebel\'ein-Renyi Maximal Correlation]
	\label{def:max_corr}
	For two random variables $Y\in\calY$ and $Z\in\calZ$, the maximal correlation between them is given by
	\begin{align}\label{eq:mc}
	\rho(Y,Z)=\sup_{f, g}{\E[f(Y)g(Z)]},
	\end{align}
	where the functions $f:\calY\mapsto\bbR$ and $g:\calZ\mapsto\bbR$ are such that
	\begin{align*}
	&\E[f(Y)]=0,\ \E[g(Z)]=0, \\
	&\E[f^2(Y)]=1,\ \E[g^2(Z)]=1.
	\end{align*}
	If the supremum in \cref{eq:mc} is attainable, the optimal $f$ and $g$ in \cref{eq:mc} are called the maximal correlation functions of $Y$ and $Z$, respectively.
\end{Definition}
Some of the well-known properties of maximal correlation are \cite{Ren:J1959}: (i) $0\leq\rho(Y,Z)\leq 1$; (ii) $\rho(Y,Z)=0$ if and only if $Y$ and $Z$ are independent; (iii) $\rho(Y,Z)=1$ if and only there exists functions $f(Y)=g(Z)$ almost surely; and (iv) $\rho(Y,Z)$ is equal to the second largest singular value of the divergence transition matrix $Q$ \cite{HuaZhe:J2014} with $Q(y,z)=\frac{p_{Y,Z}(y,z)}{\sqrt{p_{Y}(y)p_Z(z)}}$ when $\abs{\calY}$ and $\abs{\calZ}<\infty$, indicating that maximal correlation is unrelated to the symbols used for representing the random variables.

\begin{Remark}
	Given that the maximal correlation of two random variables is zero if and only if these two random variables are independent, the use of maximal correlation as a privacy function to be minimized is well motivated. However, using maximal correlation as the utility function to be maximized is questionable. For example, if the sanitized data $\tX$ is a compressed version of the raw data $X$, the maximal correlation of $\tX$ and $X$ is always equal to one (the maximum attainable). Meanwhile, it can be the case that $\tX$ leads to little or no utility.
\end{Remark}
In what follows, we show that maximal correlation can be computed by solving an optimization problem with a linear equality constraint in \cref{thm:mx_opt1} and an unconstrained optimization problem in \cref{thm:mx_opt2}. Thereafter, we propose machine learning techniques to approximate maximal correlation and use it as a privacy function. In addition, we present the kernel method as an alternative approach for maximal correlation estimation. Both methods allow us to update the sanitizer via backpropagation. 

For two random variables $Y\in\calY$ and $Z\in\calZ$ with pdfs $p_Y$ and $p_Z$ respectively, consider the following Hilbert spaces:
\begin{align*}
&\scY=\set*[\vert]{f:\calY\mapsto\bbR \given \E[f(Y)^2]<\infty},\\
&\scZ=\set*[\vert]{g:\calZ\mapsto\bbR \given \E[g(Z)^2]<\infty},
\end{align*}
with inner product defined as $\langle{f_1,f_2}\rangle_{\scY}=\E[f_1(Y)f_2(Y)]$ for $f_1,f_2\in\scY$ and $\langle{g_1,g_2}\rangle_{\scZ}=\E[g_1(Z)g_2(Z)]$ for $g_1,g_2\in\scZ$. The respective induced norms are denoted as $\norm{\cdot}_{\scY}$ and $\norm{\cdot}_{\scZ}$. 

Define a linear operator $T:\scY \mapsto \scZ$ by the conditional expectation $Tf(Z) =\E[f(Y) @| Z]$ for each $f\in\scY$. 
Assuming that
\begin{align}\label{k_assumption}
\int{\abs{k(y,z)}^2 p_Y(y) p_Z(z) \ud y \ud z}<\infty,
\end{align}
where $k(y,z)=\frac{p_{Y,Z}(y,z)}{p_Y(y) p_Z(z)}$, we conclude from \cite[Proposition 4.7]{Con:B90} that $T$ is compact. From \cite{Ren:J1959}, we have
\begin{align}\label{rho_T}
\rho(Y,Z) = \max_{\substack{f\in\scY \\ \E[f(Y)]=0}} \frac{\norm{Tf}_\scZ}{\norm{f}_\scY},
\end{align}
where we have used $\max$ instead of $\sup$ on the right-hand side to indicate that the maximal correlation is attainable. From Jensen's inequality, we have
\begin{align*}
\norm{Tf}_{\scZ}^2 = \E[\E[f(Y)]{Z}^2] \leq \E[\E[f^2(Y)]{Z}] = \norm{f}_\scY, 
\end{align*}
hence the largest singular value of $T$, $\norm{T}=1$ is attained by $f(y) = 1$ for all $y\in\calY$. Since $f\in\scY$ such that $\E[f(Y)]=0$ is orthogonal to the span of constant functions in $\scY$, we have shown that $\rho(Y,Z)$ is the second largest singular value of $T$.

\begin{Lemma} \label{lemma:opt_mc}
Assuming \cref{k_assumption}, for each $f\in\scY$, we have
\begin{align*}
\norm{Tf}_{\scZ}^2
=\max_{g\in\scZ}\braces*{2\E[f(Y)g(Z)]-\E[g^2(Z)]}.
\end{align*}
\end{Lemma}
\begin{IEEEproof}
Since $g(Z)=\E[f(Y)\mid Z]$, minimizes $\E[\left(f(Y)-g(Z)\right)^2]$ for a given $f\in\scY$, we have
	\begin{align*}
	&\min_{g\in\scZ}\E[\left(f(Y)-g(Z)\right)^2] \\
	&=\E[\left(f(Y)-\E[f(Y)\mid Z]\right)^2] \\
	&=\E[f^2(Y)]-\E[\E[f(Y)|Z]^2] \\
	&=\E[f^2(Y)]-\norm{Tf}_{\scZ}^2.
	\end{align*}
	The proof is complete by expanding the left-hand side of the above equation.
\end{IEEEproof}

\cref{lemma:opt_mc} and \cref{rho_T} give us the following result.

\begin{Proposition} \label{thm:mx_opt1}
Assuming \cref{k_assumption}, the optimal value obtained by solving
\begin{align} \label{eq:mx_opt1}
\begin{aligned}
\max_{f\in\scY,g\in\scZ} &\ \frac{2\E[f(Y)g(Z)]-\E[g^2(Z)]}{\E[f^2(Y)]} \\
\st &\ \E[f(Y)]=0.
\end{aligned}
\end{align}
is equal to $\rho^2(Y,Z)$. The optimal $f$ for \cref{eq:mx_opt1} is the maximal correlation function of $Y$.
\end{Proposition}

Supposing \cref{eq:mx_opt1} is optimized at $f=f^*$ and $g=g^*$, it can be verified that $g^*(Z)=\E[f^*(Y)\mid Z]$. Then, $\frac{g^*(Z)}{\sqrt{\E[{g^*}^2(Z)]}}$ is the maximal correlation function of $Z$, which can be seen from the Cauchy-Schwarz inequality:
\begin{align*}
\E[f(Y)g(Z)]
&=\E[g(Z)\E[f(Y)|Z]] \\
&\leq\sqrt{\E[g^2(Z)]\E[\E[f(Y)|Z]^2]},
\end{align*}
where the equality holds if and only if $g(Z)$ is proportional to $\E[f(Y)|Z]$. 

\begin{Proposition} \label{thm:mx_opt2}
Assuming \cref{k_assumption}, we have
\begin{align} 
& \max_{\mathbf{f},\bg} 
\braces*{2\trace{\E[\mathbf{f}(Y)\T\mathbf{f}(Y)]^{-1/2}\E[\mathbf{f}(Y)\T\bg(Z)]}
-\trace{\E[\bg(Z)\T\bg(Z)]}} = 1+\rho^2(Y,Z),\label{eq:mx_opt2_obj1}
\end{align}
where $\mathbf{f}(Y)=\left[f_1(Y),f_2(Y)\right]$ and $\bg(Z)=\left[g_1(Z),g_2(Z)\right]$ with $f_1,f_2\in\scY$ and $g_1,g_2\in\scZ$.
\end{Proposition}
\begin{IEEEproof}
Notice that the optimal value obtained by solving
\begin{align*}
\max_{f_1,f_2\in\scY} &\ \norm{Tf_1}_{\scZ}^2 + \norm{Tf_2}_{\scZ}^2 \\
\st &\ \norm{f_1}_{\scY}^2=1,\ \norm{f_2}_{\scY}^2=1, \\
&\ \langle{f_1,f_2}\rangle_{\scY}=0,
\end{align*}
is equal to $1+\rho^2(Y,Z)$. 
From \cref{lemma:opt_mc}, the above problem is equivalent to
\begin{align} \label{eq:mx_opt2_obj2}
\begin{aligned}
\max_{\mathbf{f},\bg} &\ 2\trace{\E[\mathbf{f}(Y)\bg(Z)\T]} - \trace{\E[\bg(Z)\T\bg(Z)]} \\
\st &\ \E[\mathbf{f}(Y)\T\mathbf{f}(Y)]=\bI,
\end{aligned}
\end{align}
where $\boldf, \bg$ are as defined in the proposition statement and $\bI$ is the identity matrix. Let $\mathbf{f}(Y)=\mathbf{f}'(Y)\bA$ where $\bA\in\bbR^{2\times2}$ is a non-singular matrix and $\bB=\E[{\mathbf{f}'(Y)}\T\mathbf{f}'(Y)]^{1/2}$. We have
\begin{align*}
\E[\mathbf{f}(Y)\T\mathbf{f}(Y)]
=\bA\T\bB\bB\bA.
\end{align*}
Since $\bA$ is non-singular and $\E[\mathbf{f}(Y)\T\mathbf{f}(Y)]=\bI$, $\bB$ is invertible. We have
\begin{align*}
\trace{\E[\mathbf{f}(Y)\bg(Z)\T]}
=\trace{\bA\T\bB\bB^{-1}\E[{\mathbf{f}'(Y)}\T\bg(Z)]}.
\end{align*}
By substituting the above equations, \cref{eq:mx_opt2_obj2} can be expressed as 
\begin{align*}
\begin{aligned}
	\max_{\mathbf{f'},\bg,\bA} &\ 2\trace{\bA\T\bB\bB^{-1}\E[{\mathbf{f}'(Y)}\T\bg(Z)]} - \trace{\E[\bg(Z)\T\bg(Z)]} \\
	\st &\ \bA\T\bB\bB\bA=\bI,
\end{aligned}
\end{align*}
which can be regarded as an orthogonal procrustes problem \cite{Sch:J1966}, whose solution leads to the left-hand side of \cref{eq:mx_opt2_obj1}. The proof is complete.
\end{IEEEproof}

\subsection{DRIP-Max}
In practice, to find the maximal correlation in \cref{eq:mc} by searching over the whole function space is infeasible. To obtain a computationally tractable implementation, we present two approaches to estimate maximal correlation by (i)~restricting maximal correlation functions to be from a rich class represented by neural networks, and (ii)~restricting maximal correlation functions to be from RKHSs. We use a minibatch set $\{(\tx_i,\bs_i)\}_{i=1}^M$  (how it is constructed is described in \cref{tab:algo}) to estimate maximal correlation between $S$ and $\tX$, and we show how to use it to form a privacy function that allows us to backpropagate through the sanitizer network. Using the same max-min problem \cref{opt:va_obj} as DRIP-Var but replacing the privacy function $\scP(\btau;\btheta)$ by the estimated maximal correlation, we obtain the DRIP-Max framework.

\subsubsection{Neural network method}
\label{subsect:nn_mx_est}

\begin{figure}[!htb]
	\centering
	\includegraphics[scale=1]{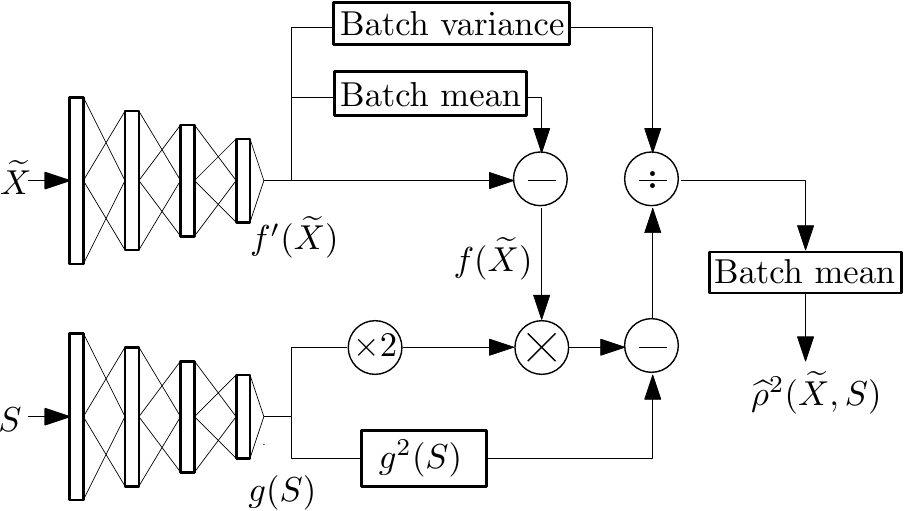}
	\caption{Neural network based maximal correlation estimator.}
	\label{fig:max_corr_net}
\end{figure}

Given a set of inputs, it is reasonable to assume that the variance of the output from the neural network is bounded and that the set of functions represented by the neural network belongs to a $\calL^2$ space. We search over the functions represented by neural networks to approximate maximal correlation. We illustrate an implementation following the constrained optimization problem in \cref{thm:mx_opt1} (using the unconstrained optimization in \cref{thm:mx_opt2} is also feasible but at the cost of more optimization parameters). Let $f'$ and $g$ represent two neural networks that take $\tX$ and $S$ as inputs, respectively. Suppose both neural networks output bounded real values. To start with, we center the data by letting
\begin{align*}
f(\tx)=f'(\tx)-\frac{1}{M}\sum_{i=1}^Mf'(\tx_i),
\end{align*}
which ensures that $\E[f(\tX)]=0$. We make the following empirical estimates: 
\begin{align*}
&\E[f^2(\tX)]\approx\frac{1}{M}\sum_{i=1}^M f^2(\tx_i), \\
&\E[g^2(S)]\approx\frac{1}{M}\sum_{i=1}^M g^2(\bs_i), \\
&\E[f(\tX)g(S)]\approx\frac{1}{M}\sum_{i=1}^M f(\tx_i)g(\bs_i).
\end{align*}
Substituting the above estimates into \cref{eq:mx_opt1} with $Y=\tX$ and $Z=S$, we obtain the estimate
\begin{align*}
\hat{\rho}^2(\tX,S)
=\max_{f',g}\frac{2\sum_{i=1}^M{f(\tx_i)g(\bs_i)}-g^2(\bs_i)}{\sum_{i=1}^M f^2(\tx_i)}.
\end{align*}
By building a machine learning model as demonstrated in \cref{fig:max_corr_net}, we can apply gradient descent to jointly optimize over $f'$ and $g$ to approximate maximal correlation.
To maintain consistency with the objective \cref{opt:va_obj} by following the format of the privacy function in \cref{eq:va_privacy}, we write the privacy function as $\scP(\btheta)=\max_{\btau}\scP(\btau;\btheta)$, where
\begin{align*}
\scP(\btau;\btheta)
=\frac{2\sum_{i=1}^M{f(\tx_i)g(\bs_i)}-g^2(\bs_i)}{\sum_{i=1}^M f^2(\tx_i)},
\end{align*}
with $\btau$ denoting the collection of the layer weights of $f'$ and $g$.

In particular, when $\abs{\calS}<\infty$, we can invoke a nonparametric estimation of the correlation function $g$ to avoid the approximation error of $g$ introduced by using neural network. Recall that $g^*(S)=\E[f^*(\tX) @| S]$ if $f^*$ and $g^*$ optimize \cref{eq:mx_opt1} with $Y=\tX$ and $Z=S$. Inspired by the alternating conditional expectations (ACE) algorithm \cite{BreFri:J1985}, we repeat the following two steps until convergence:
\begin{enumerate}[a)]
	\item Fix $g$. Update $f'$ by applying gradient descent to optimize 
	\begin{align*}
	\max_{f'}\frac{\sum_{i=1}^M f(\tx_i)g(\bs_i)}{\sum_{i=1}^M f^2(\tx_i)}.
	\end{align*}
	\item Fix $f'$. For each $\bs\in\calS$, update $g$ according to 
	\begin{align*}
	g(\bs)
	=\E[f(\tX) @| S=\bs]
	\approx\frac{1}{M}\sum_{i=1}^M f(\tx_i)\delta(\bs_i-\bs),
	\end{align*}
	where $\delta$ is the Dirac delta function.
\end{enumerate}
\subsubsection{Kernel method}
\label{subsect:ker_mx_est}
To provide a complement to the neural network method, we work with functions in RKHSs and make use of the kernel trick to search over this space efficiently \cite{HarSzeSha:J2004}. For $Y=\tX$ and $S$, let $\calH_Y$ be a RHKS of real-valued functions of $Y$, associated with the Mercer kernel $K_Y(\cdot,\cdot)$. Let $\Phi_{Y}(y)=K_{Y}(y,\cdot)$ and $\ip{\cdot}{\cdot}_Y$ be feature map and inner product, respectively. The point evaluation of $f\in\calH_Y$ can be written as $f(y)=\langle \Phi_{Y}(y), f\rangle$. The maximal correlation of $\tX$ and $S$ restricted to functions in $\calH_{\tX},\calH_S$ is
\begin{align*}
\rho_{\mathrm{K}}(\tX,S)
&=\sup_{f\in\calH_{\tX},g\in\calH_S}{\E[f(\tX)g(S)]} \\
&=\sup_{f\in\calH_{\tX},g\in\calH_S}{\E[\ip{\Phi_{\tX}(\tX)}{f} \ip{\Phi_S(S)}{g}]},
\end{align*}
with constraints $\E[f(\tX)]=0,\E[g(S)]=0$ and $\E[f^2(\tX)]=1,\E[g^2(S)]=1$.

It has been shown in \cite{BacJor:J2012} that if $\calH_\tX$ and $\calH_S$ are the RKHSs corresponding to a Gaussian kernel on $\bbR$, $\rho_{\mathrm{K}}(\tX,S)=0$ if and only if the random variables $\tX$ and $S$ are independent of each other. For $\tX$ and $S$ with finite support, \cref{lemma:ker_mx} below provides a sufficient condition for $\rho_{\mathrm{K}}(\tX,S)=\rho(\tX,S)$ under which $\tX$ and $S$ are independent if $\rho_{\mathrm{K}}(\tX,S)=0$. Therefore, the kernel method suits the cases where random variables are discrete.
\begin{Lemma} \label{lemma:ker_mx}
Suppose $\tcalX=\{\bchi_1,\ldots,\bchi_L\}$ and $\calS=\{\bnu_1,\ldots,\bnu_N\}$. Let $\bG_{\tX}$ and $\bG_S$ be the Gram matrices associated with $\tcalX$ and $\calS$, respectively. If $\rank(\bG_{\tX})=M$ and $\rank(\bG_S)=N$, $\rho_{\mathrm{K}}(\tX,S)=\rho(\tX,S)$.
\end{Lemma}
\begin{IEEEproof}
	Let $\ba=\left[a_1,\ldots,a_L\right]\in\bbR^L$ and $\bb=\left[b_1,\ldots,b_N\right]\in\bbR^N$. Let $f=\sum_{i=1}^L a_i\Phi_{\tX}(\bchi_i)$ and $g=\sum_{i=1}^N b_i\Phi_S(\bnu_i)$. Then, $f\in\calH_\tX$ and $g\in\calH_S$. Since $f(\bchi_i)=\langle{\Phi_{\tX}(\bchi_i),f}\rangle$ and $g(\bnu_i)=\langle{\Phi_S(\bnu_i),g}\rangle$, we have $\left[f(\bchi_1),\ldots,f(\bchi_L)\right]=\ba\bG_{\tX}$ and $\left[g(\bnu_1),\ldots,g(\bnu_N)\right]=\bb\bG_S$. Since $\bG_{\tX}$ and $\bG_S$ are full-rank, there exist $\ba$ and $\bb$ so that $f$ and $g$ are the maximal correlation functions of $\tX$ and $S$ given by \cref{def:max_corr}. The proof is completed by noting that $\rho_{\mathrm{K}}(\tX,S)\leq\rho(\tX,S)$.
\end{IEEEproof}

Using the minibatch data $\{(\tx_i,\bs_i)\}_{i=1}^M$, the empirical estimate of the maximal correlation of $\tX$ and $S$ is:
\begin{align} \label{opt:ker_mx}
\widehat{\rho}_{\mathrm{K}}(\tX,S)
=\frac{1}{M}\max_{f\in\calH_\tX, g\in\calH_S}{\sum_{i=1}^M{\langle{\bar{\Phi}_{\tX}(\tx_i),f}\rangle\langle{\bar{\Phi}_S(\bs_i),g}\rangle}},
\end{align}
where $\bar{\Phi}_{\tX}(\tx_i)=\Phi_{\tX}(\tx_i)-\frac{1}{M}\sum_{i=1}^M\Phi_{\tX}(\tx_i)$ and $\bar{\Phi}_S(\bs_i)=\Phi_{S}(\bs_i)-\frac{1}{M}\sum_{i=1}^M\Phi_S(\bs_i)$ are called the centered feature maps \cite{SchSmoMul:J1998} ensuring the zero mean constraint is satisfied.
Since the optimal $f$ and $g$ lie in the linear spaces spanned by the feature maps of the minibatch data, we write 
\begin{align*}
f=\sum_{i=1}^M a_i\Phi_{\tX}(\tx_i), \
g=\sum_{i=1}^M b_i\Phi_S(s_i),
\end{align*}
where $\left[a_1,\ldots,a_M\right]=\ba\T$ and $\left[b_1,\ldots,b_M\right]=\bb\T\in\bbR^M$.
Let $\bK_{\tX}$ and $\bK_S$ be the Gram matrices associated with the minibatch data points $\{\tx_i\}_{i=1}^M$ and $\{\bs_i\}_{i=1}^M$, respectively. Let $\bar{\bK}_{\tX}=\bH\bK_{\tX}\bH$ and $\bar{\bK}_S=\bH\bK_S\bH$ where $\bH=\bI-\frac{1}{M}\bone$ is a constant centering matrix ($\bI\in\bbR^{M\times M}$ is an identity matrix and $\bone\in\bbR^{M\times M}$ is an all-1s matrix). The problem becomes \cite{BacJor:J2012}:
\begin{align}\label{eq:hatrho_K}
\widehat{\rho}_{\mathrm{K}}(\tX,S)=
\max_{\ba,\bb}\frac
{\ba\T\bar{\bK}_{\tX}\bar{\bK}_S\bb}
{\sqrt{\ba\T(\bar{\bK}_{\tX}+\eta\bI)^2\ba}\sqrt{\bb\T(\bar{\bK}_S+\eta\bI)^2\bb}},
\end{align}
where $\eta$ is a small positive constant for regularization. This is equivalent to a generalized eigenvalue problem:
\begin{align} \label{eq:ker_mx_eig}
\begin{bmatrix}
\bzero && \bar{\bK}_{\tX}\bar{\bK}_S \\ 
\bar{\bK}_S\bar{\bK}_{\tX} && \bzero
\end{bmatrix}
\begin{bmatrix}
\ba \\ \bb
\end{bmatrix}
= \rho
\begin{bmatrix}
(\bar{\bK}_{\tX}+\eta\bI)^2 && \bzero \\ 
\bzero && (\bar{\bK}_S+\eta\bI)^2
\end{bmatrix}
\begin{bmatrix}
\ba \\ \bb
\end{bmatrix}.
\end{align}
Let $\ba=\ba^*$ and $\bb=\bb^*$ be the solution of \cref{eq:ker_mx_eig}. Then, \cref{opt:ker_mx} can be written as
\begin{align} \label{eq:ker_mx}
\widehat{\rho}_{\mathrm{K}}(\tX,S)
=\frac{1}{M}{\ba^*}\T\bar{\bK}_{\tX}\bar{\bK}_S\bb^*
=\frac{1}{M}\sum_{i=1}^M\sum_{j=1}^M \alpha_{i,j}K_{\tX}(\tx_i,\tx_j),
\end{align}
where $\alpha_{ij}=\left[\bH\ba^*\right]_{i}\left[\bH^2\bK_S\bH\bb^*\right]_{j}$ with $\left[\ba^*\right]_i$ being the $i$-th element of $\ba^*$.

To conform with the utility-privacy tradeoff formulation in \cref{opt:va_obj}, we take $\btau=\{\ba,\bb\}$ and $\scP(\btau;\btheta)$ to be the function on the right-hand side of \cref{eq:hatrho_K} to be maximized. Then, $\scP(\btheta)=\max_{\btau}\scP(\btau;\btheta) = \widehat{\rho}_{\mathrm{K}}(\tX,S)$. The privacy function $\scP(\btheta)$ is differentiable when the kernel $K_{\tX}$ is differentiable. For example, for any shift-invariant kernel $K_{\tX}(\tx_i,\tx_j)=K_{\tX}(\tx_i-\tx_j)$ such as the radial basis function kernel, the gradient of the privacy function has a concise expression:
\begin{align} \label{eq:ker_mc_grad}
\frac{\partial\scP(\btheta)}{\partial\btheta}
=\frac{1}{M}\sum_{i=1}^M \evalat{\frac{\partial \tX}{\partial \btheta}}{\bx_i,\bxi_i} 
\left(\sum_{j=1}^M (\alpha_{ij}-\alpha_{ji}) 
\nabla K_{\tX}(\tx_i-\tx_j)\right),
\end{align}
where $\nabla K_{\tX}(\tx)$ denotes the gradient of the kernel $K_{\tX}$ at $\tx$ and $\evalat{\frac{\partial \tX}{\partial \btheta}}{\bx_i,\bxi_i}$ is a Jacobian matrix derived from the sanitizer network at the fixed input $(\bx_i,\bxi_i)$ with $\tx_i=f^{\btheta}(\bx_i,\bxi_i)$ (cf. \cref{eq:gradient}). Thus, the gradient can be backpropagated through the sanitizer network to update the sanitization parameter $\btheta$.
\begin{Remark}
	Closely related to the kernel correlation, the paper \cite{GreBouSmo:J2005} proposed an independence criterion based on the cross-covariance operator in RKHSs \cite{FukBacJor:J2004}, which can also be used as a privacy function. The Hilbert-Schmidt Independence Criterion (HSIC) is defined as the Hilbert-Schmidt norm of the cross covariance operator:
	\begin{align*}
	\mathrm{HSIC}(\tX,S,\calH_1,\calH_2)=\norm{C_{\tX S}}_{HS}^2,
	\end{align*}
	with $C_{\tX S}=\E[\Phi_{\tX}(\tX)\otimes\Phi_S(S)]-\E[\Phi_{\tX}(\tX)]\otimes\E[\Phi_S(S)]$ where $\otimes$ denotes a tensor product \cite{Bak:J1973}.
	If $K_{\tX}$ and $K_S$ are characteristic kernels, HSIC is zero if and only if $\tX$ and $S$ are independent \cite{GreHerSmo:J2005}. The empirical estimate of HSIC results in a non-parametric independence measure between $\tX$ and $S$:
	\begin{align*}
	\widehat{\mathrm{HSIC}}(p_{\tX,S})
	&=\frac{1}{(M-1)^2}\trace{\bK_{\tX}\bH\bK_S\bH} \\
	&=\frac{1}{(M-1)^2} \sum_{i=1}^M \sum_{j=1}^M \alpha_{ij}K_{\tX}(\tx_i,\tx_j),
	\end{align*}
	where $\alpha_{ij}=\left[\bH\bK_S\bH\right]_{i,j}$. While its computational complexity $\calO(M^2)$ is less than that of the kernel correlation method where the eigendecomposition has computational complexity $\calO(M^3)$ for square matrix of size $M\times M$, experiments indicate slow convergence of the utility-privacy tradeoff when used as a privacy function. Therefore, we propose the kernel correlation approach instead.
\end{Remark}
%

%-------------------------------Section----------------------------------%
%
\section{Deep Learning Architecture for Images}\label{sect:model_dp}

In this section, we present a deep learning architecture for the proposed DRIP framework in \cref{opt:va_obj} for use with images, with the additional assumption that $\calS$ is a discrete set. We illustrate DRIP-Var using our deep learning architecture but it can be adapted to DRIP-Max as well. We assume that a training set $\{(\bx_i,\bs_i)\}_{i=1}^N$ where each $\bx_i$ is an image is available.

% sanitizer
\subsection{Sanitizer} 
\begin{figure}[!htb]
	\centering
	\includegraphics[scale=1]{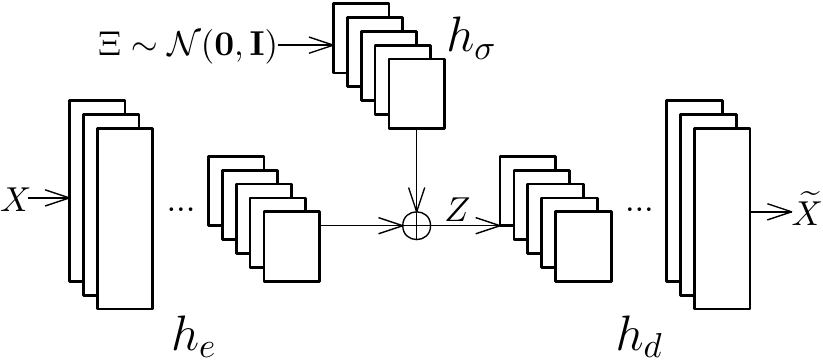}
	\caption{Probabilistic sanitizer $p^{\btheta}_{\tX|X}$.}
	\label{fig:sanitizer}
\end{figure}

We let the probabilistic sanitizer $p^{\btheta}_{\tX|X}$ be a combination of a fully-convolutional network (FCN) \cite{LonSheDar:C2015} encoder and decoder denoted as $h_e$ and $h_d$, respectively, with random noise injected in the bottleneck layers as illustrated in \cref{fig:sanitizer}.
The sanitizer compresses the raw data $X$ to a hidden representation and then perturbs it by adding noise obtained through applying a linear transformation layer $h_\sigma$ on an input Gaussian noise (the covariance of the transformed additive noise is denoted as $\bSigma$). An up-sampling decoder $h_d$ reconstructs a data from the perturbed hidden representation $Z$. This process is exactly a reparameterization trick as we introduce an auxiliary random variable $\Xi$ independent of $X$ and let $\tX$ be a function of $X$ and $\Xi$:
\begin{align*}
p_{Z|X}(\bz|\bx)=\N{h_e(\bx)}{\bSigma},\ \tX=h_d(Z).
\end{align*}
Here $\btheta$ denotes the collection of FCN weights of the encoder and decoder and the noise transformation network.

%utility function
\subsection{Utility Function \txp{$\scL(\bphi;\btheta)$}{L}} We build the utility function $\scL(\bphi;\btheta)$ in \cref{eq:va_utility} by letting $g^{\bphi}$ be a FCN with the input and output being the same shape and $\bphi$ as the collection of trainable weights. We let the variational posterior distribution of $X$ be a composition of a radial basis function and $g^{\bphi}$:
\begin{align*}
q^{\bphi}_{X|\tX}(\bx|\tx)=C\exp\left(\norm{\gamma(g^{\bphi}(\tx))-\gamma(\bx)}^2\right),
\end{align*}
where $C$ is a normalization constant and $\gamma$ is a fixed high-level feature extraction function obtained from pretrained networks (cf.\ perceptual loss in \cite{JohAlaLi:C16}). We let $\gamma$ be an identity function if no feature map function is used. 

%privacy function
\subsection{Privacy Function \txp{$\scP(\btau;\btheta)$}{P}} To form the privacy function $\scP(\btau;\btheta)$, we let the variational distribution $q^{\btau}_{S|\tX}$ be a convolutional neural network (CNN) classifier that outputs the likelihood for each element to be in $\calS$ by using the softmax activation. Let $\btau$ denote the collection of layer weights of the classifier. We rewrite \cref{eq:va_privacy} as 
\begin{align*}
\scP(\btau;\btheta)=\E_{(X,\tX)\sim p^{\btheta}_{X,\tX}}[\sum_{\bs\in\calS}p(\bs|X)\log q^{\btau}_{S|\tX}(\bs|\tX)].
\end{align*}
In order to use the training data $\{(\bx_i,\bs_i)\}_{i=1}^N$ to optimize the privacy function, we let $p(\bs_i|\bx_i)=1$ and $p(\bs|\bx_i)=0$ if $\bs\neq\bs_i$, for all $i=1,\ldots,N$ (hard labels). Optimizing $\scP(\btau;\btheta)$ over $\btau$ is similar to the adversarial training in GAN \cite{GooAbaMir:C2014}.

%regularizer
\subsection{Regularization \txp{$\scR(D_X;\btheta)$}{R}} 
\begin{figure}[!htb]
\centering
\begin{subfigure}{.38\textwidth}
\centering
\includegraphics[scale=0.8]{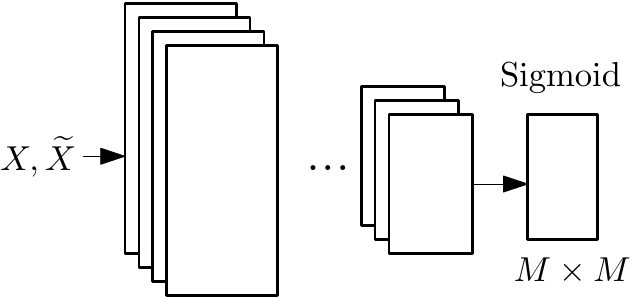}
\caption{Patch discriminator}\label{fig:patch_dis}
\end{subfigure}
\begin{subfigure}{.30\textwidth}
\centering
\includegraphics[scale=0.8]{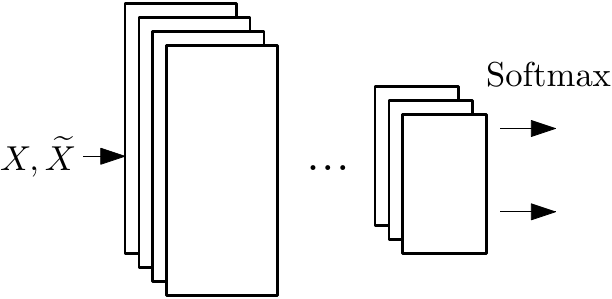}
\caption{Normal discriminator}\label{fig:normal_dis}
\end{subfigure}
\caption{Different types of discriminator.}
\label{fig:discriminator}
\end{figure}

We use domain adaption as the regularizer $\scR(D_X;\btheta)$ since it has been demonstrated to perform well in image to image translation \cite{ZhuPar:J2018}. Denote $D_X$ as the discriminator that seeks to distinguish $X$ from $\tX$. Let $D_X$ be a FCN with $M\times M$ outputs where the sigmoid activation function is applied on the last layer to convert numerical values to probabilities as illustrated in \cref{fig:patch_dis}. Each output represents how likely the corresponding patch of an input image is sampled from $p_X$. This structure is originated from the patch discriminator \cite{IsoZhu:J2018}. Compared to a classical normal discriminator shown in \cref{fig:normal_dis} that outputs the likelihood of a binary hypothesis, the patch discriminator is capable of capturing both the high and low frequency features of the input image and is thus more robust. Denote $D_X^{(i)}$ as the output corresponding to the $i$-th patch. By averaging the discriminator loss over all the patches, we obtain a modified version of \cref{eq:rgl_da} as follows:
\begin{align}\label{mod_scR}
\begin{split}
\scR(D_X;\btheta)=\sum_{i=1}^{M^2}\E_{X\sim p_X}[\log D_X^{(i)}(X)] 
+\sum_{i=1}^{M^2}\E_{\tX\sim p_{\tX}}[\log\parens*{1-\sum_{i=1}^{M^2}D_X^{(i)}(\tX)}].
\end{split}
\end{align}

\subsection{Alternating Optimization} We propose an alternating optimization method to optimize \cref{opt:va_obj}. We first sample an initial minibatch data $\{(\bx_{j_i},\bs_{j_i})\}_{i=1}^M$ from the training set. We draw $M$ \gls{iid} noise samples from the standard normal distribution and feed the noise samples and $\{\bx_{j_i}\}_{i=1}^M$ into the sanitizer in \cref{fig:sanitizer} to obtain $\{\tx_{j_i}\}_{i=1}^M$. We use $\{(\tx_{j_i},\bx_{j_i},\bs_{j_i})\}_{i=1}^M$ to form Monte Carlo estimates of \cref{eq:va_utility,eq:va_privacy,mod_scR}, respectively, since they are written as an expectation \gls{wrt} $X$, $\tX$ and $S$. Next, we freeze the sanitizer's parameter $\btheta$ and solve \cref{utility,va_approx,va_rgl} separately. We update $\btheta$ by maximizing \cref{opt:va_obj} with $\bphi$, $\btau$ and $D_X$ being fixed. We repeat these three steps until an equilibrium is reached. A minibatch stochastic descent algorithm is outlined in \cref{tab:algo} for training the DRIP-Var framework. Note that \cref{tab:algo} can also be applied for optimizing \cref{opt:va_obj} using the maximal correlation privacy functions under the DRIP-Max framework.
\begin{algorithm}[!htb]
\caption{Minibatch stochastic gradient algorithm}\label{tab:algo}
\begin{algorithmic}[1]
\STATE{Initialize {$\btheta,\bphi,\btau,D_X$}}.
\REPEAT
\STATE{Draw a mini-batch data points $\{(\bx_{j_i},\bs_{j_i})\}_{i=1}^M$ from the training dataset and $M$ \gls{iid} noise samples $\{\bxi_i\}_{i=1}^M$ from $\calN(\bzero,\bI)$. Compute $\tx_{j_i}=h_d(h_e(\bx_{j_i})+h_{\sigma}(\bxi_i))$ to obtain $\{(\tx_{j_i},\bx_{j_i},\bs_{j_i})\}_{i=1}^M$.}
\STATE{Estimate \cref{eq:va_utility,eq:va_privacy,mod_scR} by using the $M$ data points $\{(\tx_{j_i},\bx_{j_i},\bs_{j_i})\}_{i=1}^M$.}
\STATE{Update $\bphi,\btau,D_X$ by applying stochastic gradient descent to optimize \cref{opt:va_obj}.}
\STATE{Update $\btheta$ by applying stochastic gradient descent to optimize \cref{opt:va_obj}.}
\UNTIL{$\btheta$ converges}
\end{algorithmic}
\end{algorithm}
%

%-------------------------------Section----------------------------------%
\section{Numerical Experiments}\label{sect:experiment}

In this section, we evaluate our proposed DRIP frameworks on various data sets and compare our method with state-of-the-art benchmark algorithms. 

\subsection{DRIP-Var}

We evaluate the deep learning model for DRIP-Var sketched in \cref{sect:model_dp} by testing it on the UTKFace dataset \cite{ZhaSonQi:C2017}, which contains over $20,000$ face images of size $64\times 64$ with annotations of gender, age and ethnicity. We select a subset of $7,307$ images, consisting of youngsters in the age group of $0$ to $20$, and elders in the age group of $50$ to $70$. Each age generation accounts for around half of the population and among each generation, around half of the population is female and the other half is male. Among these $7,307$ images, $5,846$ images are used for the training and $1,461$ images are used for evaluation. We convert the RGB images to grayscale. We let gender be the private variable $S$ and test the age group prediction accuracy of the sanitized images on a pretrained model (legacy system).

\subsubsection{Setup}
The setup of the deep learning model in \cref{sect:model_dp} is given in \cref{tab:drip_var_configs}, where the residual block (Res-BLK) \cite{KaiXiaSha:J2015} is made of $2$ layers of Conv2D with kernel size $4\times 4$, stride $1$ and the number of filters same with the layer it connected to. %
%The probabilistic sanitizer is a FCN with transposed convolutions \cite{MasJonMei:C2011} and its architecture is delineated in \cref{tab:drip_max_var_sanitizer}, where the residual block (Res-BLK) \cite{KaiXiaSha:J2015} is made of $2$ layers of Conv2D with kernel size $4\times 4$, stride $1$ and filter size same with the layer it connected to. The architecture of the domain adaption discriminator $D_X$ is depicted in \cref{tab:drip_var_ds} while \cref{tab:drip_var_q} describes the architecture of the variational distribution $q^{\btau}_{S|\tX}$. The architecture of the function $g^{\bphi}$ for utility is outlined in \cref{tab:drip_var_g}. 
We apply batch normalization and LeakyReLU activation with $\alpha=0.2$ on all the convolutional layers but the last layer of each model where Sigmoid, Sigmoid, Softmax and Sigmoid activation are used respectively, set $\lambda_1=0.25,\lambda_2=0.5$ and use the Adam optimizer with learning rate $0.0001$ and momentum $0.5$. We train the model by following the steps depicted in \cref{tab:algo} until an equilibrium point is obtained. Moreover, we train an age group classifier on the raw images using the ResNet-18 \cite{KaiXiaSha:J2015}, which serves as the legacy system and is used to test the utility of the sanitized images. After obtaining the sanitized images, we train a gender classifier on the sanitized images using the ResNet-18 acting as the adversary.

\begin{table}[!htbp]
	\centering
	\caption{DRIP-Var deep learning model}
	\label{tab:drip_var_configs}
	\vspace{-10pt}
	\begin{subtable}{1.0\linewidth}
	\centering
	\caption{Sanitizer $p_{\tX|X}^{\btheta}$}
	\label{tab:drip_max_var_sanitizer}
	\begin{tabular}{c|c|c|c|c} 
		\hline
		Layer & Output & Kernel size & Stride & Filters \\
		\hline
		Input       & $64 \times 64 \times 1$   &              &     &      \\ 
		Conv2D      & $64 \times 64 \times 32$  & $6 \times 6$ & $1$ & $32$ \\
		Conv2D      & $32 \times 32 \times 64$  & $3 \times 3$ & $2$ & $64$ \\
		Conv2D      & $16 \times 16 \times 128$ & $3 \times 3$ & $2$ & $128$\\ 
		Res-BLK     & $16 \times 16 \times 128$ &              &     &      \\
		\hline
		Input       & $16 \times 16 \times 1$   &              &     &      \\ 
		Conv2D      & $16 \times 16 \times 64$  & $4 \times 4$ & $1$ & $64$ \\
		Conv2D      & $16 \times 16 \times 128$ & $4 \times 4$ & $1$ & $128$\\ 
		\hline	
		Add         & $16 \times 16 \times 128$ &              &     &      \\
		Res-BLK     & $16 \times 16 \times 128$ &              &     &      \\
		Res-BLK     & $16 \times 16 \times 128$ &              &     &      \\
		Conv2DTrans & $32 \times 32 \times 64$  & $3 \times 3$ & $2$ & $64$ \\ 
		Conv2DTrans & $64 \times 64 \times 1$   & $3 \times 3$ & $2$ & $1$  \\
		\hline
	\end{tabular}
	\end{subtable}
	
	\begin{subtable}{0.9\linewidth}
	\centering
	\caption{Domain adaption discriminator $D_X$}
	\label{tab:drip_var_ds}
	\begin{tabular}{c|c|c|c|c} 
		\hline
		Layer & Output & Kernel size & Stride & Filters \\
		\hline
		Input  & $64 \times 64 \times 1$   &              &     &      \\ 
		Conv2D & $32 \times 32 \times 32$  & $4 \times 4$ & $2$ & $32$ \\
		Conv2D & $16 \times 16 \times 64$  & $4 \times 4$ & $2$ & $64$ \\ 
		Conv2D & $8  \times 8  \times 128$ & $4 \times 4$ & $2$ & $128$\\
		Conv2D & $8  \times 8  \times 32$  & $4 \times 4$ & $1$ & $32$ \\ 
		Conv2D & $8  \times 8  \times 1$   & $4 \times 4$ & $1$ & $1$  \\
		\hline
	\end{tabular}
	\end{subtable}

	\begin{subtable}{0.9\linewidth}
	\centering
	\caption{Variational distribution $q^{\btau}_{S|\tX}$}
	\label{tab:drip_var_q}
	\begin{tabular}{c|c|c|c|c} 
		\hline
		Layer & Output & Kernel size & Stride & Filters \\
		\hline
		Input  & $64 \times 64 \times 1$   &              &     &      \\ 
		Conv2D & $32 \times 32 \times 32$  & $4 \times 4$ & $2$ & $32$ \\
		Conv2D & $16 \times 16 \times 64$  & $4 \times 4$ & $2$ & $64$ \\ 
		Conv2D & $8  \times 8  \times 128$ & $4 \times 4$ & $2$ & $128$\\ 
		Conv2D & $8  \times 8  \times 32$  & $4 \times 4$ & $1$ & $32$ \\ 
		Dense  & $2$                       &              &     &      \\
		\hline
	\end{tabular}
	\end{subtable}

	\begin{subtable}{0.9\linewidth}
	\centering
	\caption{$g^\phi$ for variational posterior $q^{\bphi}_{X|\tX}$}
	\label{tab:drip_var_g}
	\begin{tabular}{c|c|c|c|c} 
		\hline
		Layer & Output & Kernel size & Stride & Filters \\
		\hline
		Input  & $64 \times 64 \times 1$  &              &     &     \\ 
		Conv2D & $64 \times 64 \times 32$ & $6 \times 6$ & $1$ & $32$\\
		Conv2D & $64 \times 64 \times 64$ & $3 \times 3$ & $1$ & $64$\\ 
		Conv2D & $64 \times 64 \times 32$ & $3 \times 3$ & $1$ & $32$\\
		Conv2D & $64 \times 64 \times 1$  & $3 \times 3$ & $1$ & $1$ \\
		\hline
	\end{tabular}
	\end{subtable}
\end{table}

\subsubsection{Results}
\begin{figure}[!htb]
	\centering
	\includegraphics[scale=0.4]{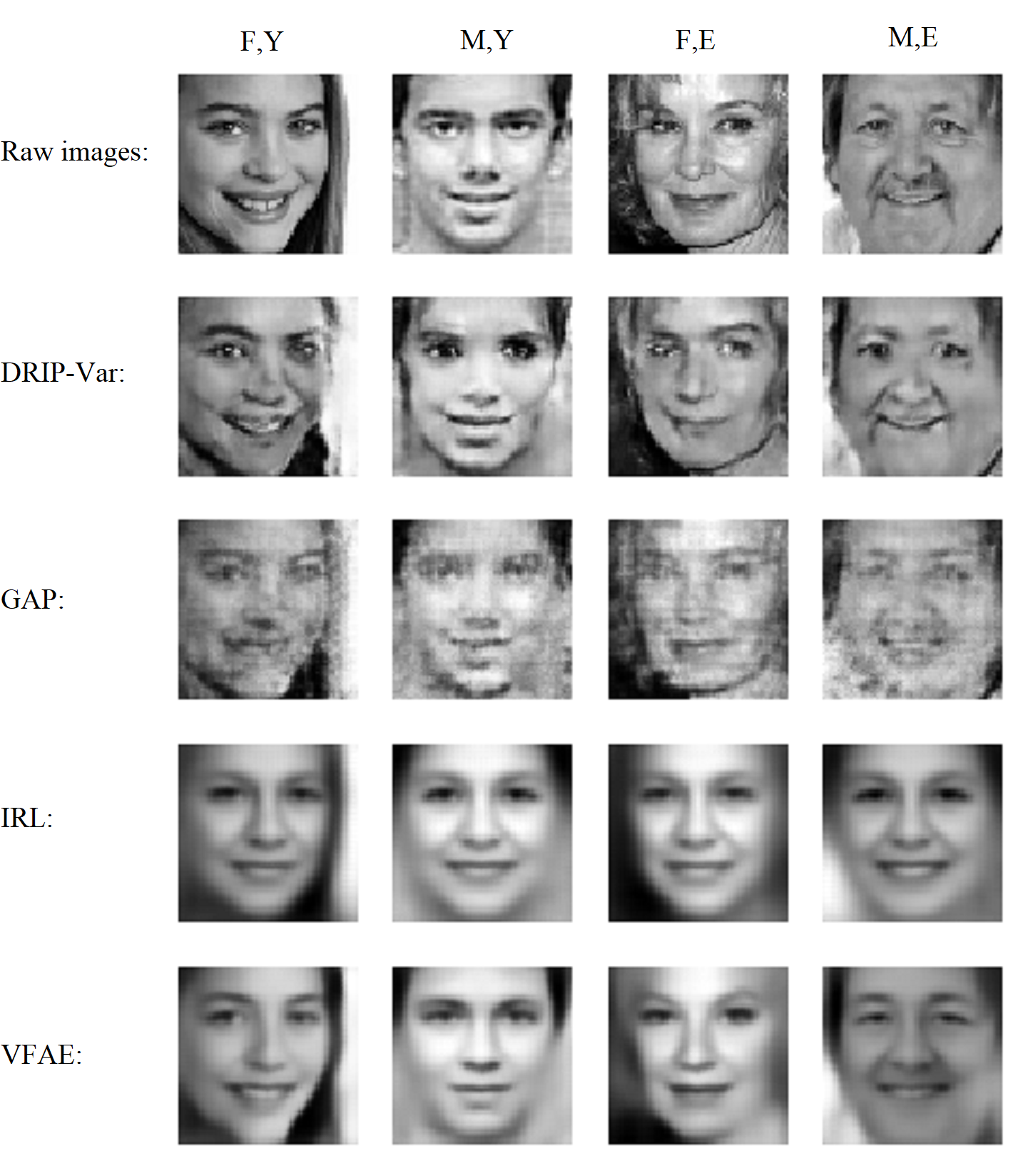}
	\caption{The true labels are displayed on top, where M, F, Y, and E denote male, female, youngster, and elder, respectively. The 1st row shows the original images. Gender is the private variable and age group classification is the utility.}
	\label{fig:drip_var_results}
\end{figure}

After the training is done, we feed the raw images to the sanitizer to obtain sanitized images. A few samples of the raw images and sanitized images are shown in the first two rows of \cref{fig:drip_var_results}, respectively. It can be observed that the gender information is obfuscated while other facial features are preserved (e.g., it is easy to match each sanitized image to its corresponding raw image without any prior knowledge). The gender accuracy of the sanitized images for the adversary is $61\%$ (compared to $50\%$ for perfect privacy). To test the utility, we input the sanitized images to the pre-trained ResNet-18 age group classifier. We obtained an accuracy of $89\%$. This accuracy can be further improved to $95\%$ if we include the age group classifier loss on the sanitized images into our objective function in training (cf.\ \cref{rem:public_variable}).

We compare the proposed DRIP-Var with the state-of-the-art GAP \cite{ChoPetXia:J2018}, invariant representations learning (IRL) \cite{DanShuRob:J2018}, and variational fair autoencoder (VFAE) \cite{ChrKevYuj:J2015}. Given a raw data $\bx$ and sensitive variable $\bs$, both IRL and VFAE learn an encoder and decoder pair $q(\bz|\bx)$ and $p(\bx|\bs,\bz)$ to resolve the dependencies between the encoded hidden vector $\bz$ and the sensitive variable $\bs$. We reconstruct the sanitized image $\tx$ from $\bz$ using a decoder by minimizing the pixel distortion between $\tx$ and the raw image $\bx$. Some samples are shown in \cref{fig:drip_var_results}.

As shown in \cref{tab:drip_var_results}, DIRP-Var achieves the best utility-privacy tradeoff: it achieves higher age group classification accuracy but lower gender classification accuracy than the other methods. In addition, we observe more visual distortions in the sanitized images produced by GAP. This is due to the domain adaption technique incorporated in DRIP-Var while GAP only considers the pixel distortion as the utility. Moreover, unlike the sanitizer in GAP which contains several fully connected layers, our proposed sanitizer is lightweight since it is a FCN, which enables it to cope with high dimensional inputs and makes it suitable for end-user devices with limited resources and computing power. It can also be observed from \cref{fig:drip_var_results} that IRL is better than VFAE in terms of removing the gender information. However, both IRL and VFAE do not retain much details of the raw images. This may be because the normal distribution of the hidden vector $\bz$ imposed by these methods forces the encoder to extract only the general features of the images.

\begin{table}[!htb]
	\centering
	\caption{Image classification results. Gender is the private variable.}
	\label{tab:drip_var_results}
	\begin{tabular}{|l|c|c|} 
		\hline
		Method & Gender accuracy & Age group accuracy \\
		\hline\hline
		ResNet-18 (raw images) & 99.6\% & 99.1\% \\
		\hline
		DRIP-Var & 61.0\% & 89.0\% \\
		\hline
		GAP      & 61.5\% & 70.0\% \\
		\hline
		IRL      & 76.0\% & 57.0\% \\
		\hline
		VFAE     & 78.0\% & 58.0\% \\
		\hline
	\end{tabular}
\end{table}

\subsection{DRIP-Max}
\label{sect:results_max_corr}

We evaluate the use of maximal correlation as the privacy function described in \cref{sect:max_corr} on the German credit dataset that was previously used by \cite{ZemWu:C2013} and obtained from the UCI machine learning repository \cite{DuaGra:2019}. The German credit dataset contains $1000$ samples of personal financial data and the objective is to predict whether a person has a good credit score based on $20$ qualitative and quantitative observations such as credit amount and employment status. The entries for each numerical variable are scaled in range of $\left[0,1\right]$. The whole training set is split into $800$ training data and $200$ testing data. We choose the age (continuous variable) to be the private variable and the credit score (binary variable) to be public variable in Case 1, and reverse the choices in Case 2. The rest of the $19$ observations are treated as raw data.

\subsubsection{Setup}
In Case 1, we construct a privacy function using the neural network method depicted in \cref{subsect:nn_mx_est}. We redefine the utility function $\calL(\bphi;\btheta)$ in \cref{opt:va_obj} to be the negative cross-entropy loss of a gender classifier where $\bphi$ denotes the trainable weights of the classifier.

In Case 2, we construct a privacy function using the kernel method depicted in \cref{subsect:ker_mx_est} where we set $\eta=0.01$ and $K_{\tX}$ and $K_S$ to be the RBF kernel with bandwidth $\sigma=1$. We redefine the utility function $\calL(\bphi;\btheta)$ in \cref{opt:va_obj} to be the negative mean absolute error (MAE) loss of an age regressor where $\bphi$ denotes the trainable weights of the regressor.

\begin{figure}[!htb]
	\centering
	\begin{subfigure}{.30\textwidth}
		\centering
		\includegraphics[scale=1]{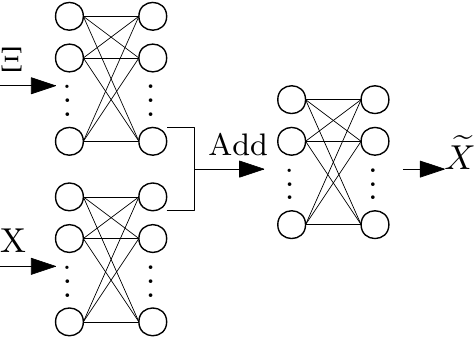}
		\caption{Sanitizer}
	\end{subfigure}
	\begin{subfigure}{.35\textwidth}
		\centering
		\includegraphics[scale=1]{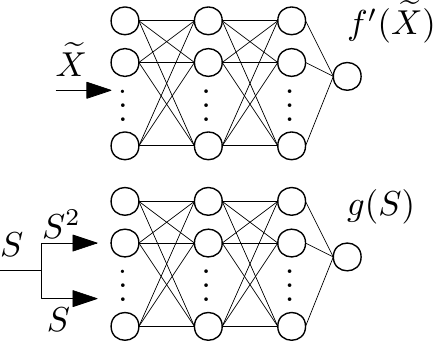}
		\caption{Maximal correlation functions}
	\end{subfigure}
	
	\begin{subfigure}{.30\textwidth}
		\centering
		\includegraphics[scale=1]{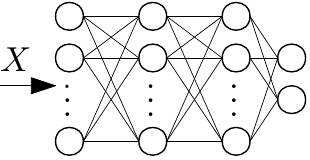}
		\caption{Credit score classifier}
	\end{subfigure}
	\begin{subfigure}{.30\textwidth}
		\centering
		\includegraphics[scale=1]{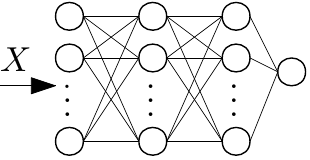}
		\caption{Age regressor}
	\end{subfigure}
	\caption{DRIP-Max configurations.}
	\label{fig:drip_max_configs}
\end{figure}

The configurations of the sanitizer, maximal correlation function networks ($f'$ and $g$ illustrated in \cref{fig:max_corr_net} in reference to \cref{subsect:nn_mx_est}), age regressor and credit score classifier are illustrated in \cref{fig:drip_max_configs}, where dense layers with $20$ units and batch normalization are employed followed by LeakyReLU activation with $\alpha=0.1$, and Sigmoid activation, Linear activation, Softmax activation, Sigmoid activation, are used for each model respectively. The age regressor and credit score classifier serve as the adversary and utility models.

We use MMD as the regularization term in \cref{opt:va_obj} with $\lambda_2=0.1$ in this experiment and we use the Adam optimizer with learning rate $0.0001$ to train the model following \cref{tab:algo}.

\subsubsection{Results}

To evaluate the sanitizer, we train an age regressor and credit score classifier on the sanitized data to obtain the adversarial loss for Cases 1 and 2, respectively. We compare against two prominent tools in compressive privacy: the discriminant component analysis (DCA) for Case 1 and the desensitized principal component analysis (DPCA) for Case 2, as well as the aforementioned VFAE method for both the cases. DCA and DPCA proposed by \cite{Kung:J2017b} perform data compression using the concept of the different utility and privacy signal subspaces.

In \cref{tab:nn_put}, the 2nd column shows the maximal correlation between the age and sanitized data, the 3rd column shows the adversarial loss on the age regression, and the 4th and 5th columns show the cross-entropy loss and accuracy of the credit score classification (utility model). A down arrow $\downarrow$ indicates that a smaller value is better for utility-privacy tradeoff. An up arrow $\uparrow$ indicates the opposite. Same notations are applied to \cref{tab:ker_put}. It can be observed that a larger $\lambda_1$ leads to greater privacy at the cost of lower utility. A lower maximal correlation leads to a larger adversarial loss. Thus, the proposed maximal correlation estimators seem to be a good privacy measure. It can be observed that DRIP-Max outshines DPCA and VFAE while DCA is comparable with DRIP-Max in terms of utility-privacy tradeoff. However, unlike DRIP-Max, DCA and DPCA cannot adjust the utility-privacy tradeoff and can only handle categorical utility and private variables, thus have only limited applications.

\begin{table}[!htbp]
\centering
\caption{German credit data results}
\label{tab:drip_max_results}
\vspace{-10pt}
\begin{subtable}{1.0\linewidth}
\centering
\caption{Case 1: Age is private}
\label{tab:nn_put}
\begin{tabular}{|l|c|c|c|c|} 
\hline
Method & Age max. corr. $(\downarrow)$  & Age MAE $(\uparrow)$ & Credit score loss $(\downarrow)$ & Credit score accuracy $(\uparrow)$\\
\hline\hline
Raw data                & 0.85 & 0.14 & 0.52 & 78\%\\ 
\hline
DRIP-Max($\lambda_1=1$) & 0.79 & 0.21 & 0.56 & 77.2\%\\ 
\hline
DRIP-Max($\lambda_1=4$) & 0.58 & 0.33 & 0.61 & 71.1\%\\ 
\hline
VFAE                    & 0.81 & 0.18 & 0.62 & 72\%\\ 
\hline
DCA                     & 0.56 & 0.34 & 0.65 & 69.9\%\\ 
\hline
\end{tabular}
\end{subtable}
\begin{subtable}{1.0\linewidth}
\centering
\caption{Case 2: Credit score is private}
\label{tab:ker_put}
\begin{tabular}{|l|c|c|c|c|} 
\hline
Method  & Credit score loss $(\uparrow)$ & Credit score accuracy $(\downarrow)$ & Age max. corr. $(\uparrow)$ & Age MAE $(\downarrow)$\\
\hline\hline
Raw data                & 0.52 & 78\%   & 0.85 & 0.14\\ 
\hline
DRIP-Max($\lambda_1=1$) & 0.61 & 73.2\% & 0.83 & 0.15\\ 
\hline
DRIP-Max($\lambda_1=2$) & 0.63 & 69.7\% & 0.82 & 0.16\\ 
\hline
VFAE                    & 0.60 & 74.6\%   & 0.82 & 0.15\\ 
\hline
DPCA                    & 0.59 & 76\%   & 0.81 & 0.16\\ 
\hline
\end{tabular}
\end{subtable}
\end{table}
%

%-------------------------------Section----------------------------------%
\section{Conclusion}
\label{sect:conclusion}
We have proposed a data-driven privacy-preserving framework that produces sanitized data compatible with legacy inference systems and which prevent leakage of sensitive private information. We presented a probabilistic sanitizer and proposed a variational method to form the privacy and utility metrics as well as an alternative privacy metric based on maximal correlation. We introduced domain regularization techniques to ensure legacy compatibility of the sanitized data. Numerical results demonstrate that our approach has better utility-privacy tradeoff than current state-of-the-art methods. In future research, it would be of interest to test the framework on more real world issues.

\bibliographystyle{IEEEtran}
\bibliography{IEEEabrv,StringDefinitions,BibBooks,refs}

\end{document}

%% file: preamble.tex
\usepackage[T1]{fontenc}
\usepackage{amsmath,amssymb,amsfonts,mathrsfs,bm}% Typical maths resource packages
\usepackage{mathtools}
\usepackage{amsthm}
\usepackage{nicefrac}
\usepackage{cite}
\usepackage[shortlabels]{enumitem}
\usepackage{graphicx}
\usepackage{epstopdf}
\usepackage{url}
\usepackage{colortbl}
\usepackage{booktabs}
\usepackage{multirow}
\usepackage[table,dvipsnames]{xcolor}
\usepackage[normalem]{ulem}

\usepackage{array}
\newcolumntype{L}[1]{>{\raggedright\let\newline\\\arraybackslash\hspace{0pt}}m{#1}}
\newcolumntype{C}[1]{>{\centering\let\newline\\\arraybackslash\hspace{0pt}}m{#1}}
\newcolumntype{R}[1]{>{\raggedleft\let\newline\\\arraybackslash\hspace{0pt}}m{#1}}

\makeatletter
\let\MYcaption\@makecaption
\makeatother
\usepackage[font=footnotesize]{subcaption}
\makeatletter
\let\@makecaption\MYcaption
\makeatother

\usepackage{xparse}



\usepackage{glossaries}

\makeatletter
\let\oldgls\gls
\let\oldglspl\glspl

\newcommand\fussy@ifnextchar[3]{%
  \let\reserved@d=#1%
  \def\reserved@a{#2}%
  \def\reserved@b{#3}%
  \futurelet\@let@token\fussy@ifnch}
\def\fussy@ifnch{%
  \ifx\@let@token\reserved@d
    \let\reserved@c\reserved@a 
  \else
    \let\reserved@c\reserved@b
  \fi
 \reserved@c}

\renewcommand{\gls}[1]{%
  \oldgls{#1}\fussy@ifnextchar.{\@checkperiod}{\@}}
\renewcommand{\glspl}[1]{%
  \oldglspl{#1}\fussy@ifnextchar.{\@checkperiod}{\@}}

\newcommand{\@checkperiod}[1]{%
  \ifnum\sfcode`\.=\spacefactor\else#1\fi
}
\makeatother

\newacronym{wrt}{w.r.t.}{with respect to}
\newacronym{RHS}{RHS}{right-hand side}
\newacronym{LHS}{LHS}{left-hand side}
\newacronym{iid}{i.i.d.}{independent and identically distributed}
\usepackage{float}

\ifx\notloadhyperref\undefined
	\ifx\loadbibentry\undefined
		\usepackage[hidelinks,hypertexnames=false]{hyperref} 
	\else
		\usepackage{bibentry}
		\makeatletter\let\saved@bibitem\@bibitem\makeatother
		\usepackage[hidelinks,hypertexnames=false]{hyperref}
		\makeatletter\let\@bibitem\saved@bibitem\makeatother
	\fi
\else
	\ifx\loadbibentry\undefined
		\relax
	\else
		\usepackage{bibentry}
	\fi
\fi

\usepackage[capitalize]{cleveref}
\crefname{equation}{}{}
\Crefname{equation}{}{}
\crefname{claim}{claim}{claims}
\crefname{step}{step}{steps}
\crefname{line}{line}{lines}
\crefname{condition}{condition}{conditions}
\crefname{dmath}{}{}
\crefname{dseries}{}{}
\crefname{dgroup}{}{}

\crefname{Problem}{Problem}{Problems}
\crefformat{Problem}{Problem~(#2#1#3)}
\crefrangeformat{Problem}{Problems~(#3#1#4) to~(#5#2#6)}

\crefname{Theorem}{Theorem}{Theorems}
\crefname{Corollary}{Corollary}{Corollaries}
\crefname{Proposition}{Proposition}{Propositions}
\crefname{Lemma}{Lemma}{Lemmas}
\crefname{Definition}{Definition}{Definitions}
\crefname{Example}{Example}{Examples}
\crefname{Assumption}{Assumption}{Assumptions}
\crefname{Remark}{Remark}{Remarks}
\crefname{Rem}{Remark}{Remarks}
\crefname{remarks}{Remarks}{Remarks}
\crefname{Appendix}{Appendix}{Appendices}
\crefname{Exercise}{Exercise}{Exercises}
\crefname{Theorem_A}{Theorem}{Theorems}
\crefname{Corollary_A}{Corollary}{Corollaries}
\crefname{Proposition_A}{Proposition}{Propositions}
\crefname{Lemma_A}{Lemma}{Lemmas}
\crefname{Definition_A}{Definition}{Definitions}

\usepackage{crossreftools}
\ifx\notloadhyperref\undefined
\else
	\relax
\fi

\usepackage{algorithm,algorithmic}

\ifx\loadbreqn\undefined
	\relax
\else
	\usepackage{breqn} 
\fi

\ifx\renewtheorem\undefined
\ifx\useTheoremCounter\undefined
\newtheorem{Theorem}{Theorem}
\newtheorem{Corollary}{Corollary}
\newtheorem{Proposition}{Proposition}
\newtheorem{Lemma}{Lemma}
\else

\newtheorem{Proposition}[theorem]{Proposition}
\fi

\newtheorem{Definition}{Definition}

\newtheorem{Remark}{Remark}

\fi

\newcommand{\nn}{\nonumber\\}

\theoremstyle{remark}

\theoremstyle{plain}

\newcommand{\qednew}{\nobreak \ifvmode \relax \else
      \ifdim\lastskip<1.5em \hskip-\lastskip
      \hskip1.5em plus0em minus0.5em \fi \nobreak
      \vrule height0.75em width0.5em depth0.25em\fi}

\newcommand{\calA}{\mathcal{A}}
\newcommand{\calB}{\mathcal{B}}

\newcommand{\calD}{\mathcal{D}}

\newcommand{\calF}{\mathcal{F}}

\newcommand{\calH}{\mathcal{H}}

\newcommand{\calL}{\mathcal{L}}

\newcommand{\calN}{\mathcal{N}}
\newcommand{\calO}{\mathcal{O}}

\newcommand{\calS}{\mathcal{S}}

\newcommand{\calX}{\mathcal{X}}
\newcommand{\calY}{\mathcal{Y}}
\newcommand{\calZ}{\mathcal{Z}}

\newcommand{\ba}{\mathbf{a}}
\newcommand{\bA}{\mathbf{A}}
\newcommand{\bb}{\mathbf{b}}
\newcommand{\bB}{\mathbf{B}}

\newcommand{\boldf}{\mathbf{f}}

\newcommand{\bg}{\mathbf{g}}
\newcommand{\bG}{\mathbf{G}}

\newcommand{\bH}{\mathbf{H}}

\newcommand{\bI}{\mathbf{I}}

\newcommand{\bK}{\mathbf{K}}

\newcommand{\bs}{\mathbf{s}}

\newcommand{\bx}{\mathbf{x}}

\newcommand{\bz}{\mathbf{z}}

\newcommand{\bbR}{\mathbb{R}}

\newcommand{\scL}{\mathscr{L}}

\newcommand{\scP}{\mathscr{P}}

\newcommand{\scR}{\mathscr{R}}

\newcommand{\scY}{\mathscr{Y}}
\newcommand{\scZ}{\mathscr{Z}}

\DeclareSymbolFont{bsfletters}{OT1}{cmss}{bx}{n}
\DeclareSymbolFont{ssfletters}{OT1}{cmss}{m}{n}
\DeclareMathSymbol{\bsfGamma}{0}{bsfletters}{'000}
\DeclareMathSymbol{\ssfGamma}{0}{ssfletters}{'000}
\DeclareMathSymbol{\bsfDelta}{0}{bsfletters}{'001}
\DeclareMathSymbol{\ssfDelta}{0}{ssfletters}{'001}
\DeclareMathSymbol{\bsfTheta}{0}{bsfletters}{'002}
\DeclareMathSymbol{\ssfTheta}{0}{ssfletters}{'002}
\DeclareMathSymbol{\bsfLambda}{0}{bsfletters}{'003}
\DeclareMathSymbol{\ssfLambda}{0}{ssfletters}{'003}
\DeclareMathSymbol{\bsfXi}{0}{bsfletters}{'004}
\DeclareMathSymbol{\ssfXi}{0}{ssfletters}{'004}
\DeclareMathSymbol{\bsfPi}{0}{bsfletters}{'005}
\DeclareMathSymbol{\ssfPi}{0}{ssfletters}{'005}
\DeclareMathSymbol{\bsfSigma}{0}{bsfletters}{'006}
\DeclareMathSymbol{\ssfSigma}{0}{ssfletters}{'006}
\DeclareMathSymbol{\bsfUpsilon}{0}{bsfletters}{'007}
\DeclareMathSymbol{\ssfUpsilon}{0}{ssfletters}{'007}
\DeclareMathSymbol{\bsfPhi}{0}{bsfletters}{'010}
\DeclareMathSymbol{\ssfPhi}{0}{ssfletters}{'010}
\DeclareMathSymbol{\bsfPsi}{0}{bsfletters}{'011}
\DeclareMathSymbol{\ssfPsi}{0}{ssfletters}{'011}
\DeclareMathSymbol{\bsfOmega}{0}{bsfletters}{'012}
\DeclareMathSymbol{\ssfOmega}{0}{ssfletters}{'012}

\newcommand{\btheta}{\bm{\theta}}

\newcommand{\bnu}{\bm{\nu}}
\newcommand{\btau}{\bm{\tau}}

\newcommand{\bchi}{\bm{\chi}}
\newcommand{\bphi}{\bm{\phi}}

\newcommand{\bxi}{\bm{\xi}}

\newcommand{\bSigma	}{\bm{\Sigma}}

\DeclareMathOperator{\st}{s.t.}

\DeclareMathOperator{\Tr}{Tr}

\DeclareMathOperator{\rank}{rank}

\DeclarePairedDelimiter\abs{\lvert}{\rvert}
\DeclarePairedDelimiter\parens{(}{)}

\DeclarePairedDelimiter\braces{\{}{\}}

\DeclarePairedDelimiterX\ip[2]{\langle}{\rangle}{#1,#2}
\DeclarePairedDelimiterX\norm[1]{\lVert}{\rVert}{#1}
\DeclarePairedDelimiterXPP\col[1]{\operatorname{col}}{\{}{\}}{}{#1} % column vector
\DeclarePairedDelimiterXPP\row[1]{\operatorname{row}}{\{}{\}}{}{#1} % row vector
\DeclarePairedDelimiterXPP\erf[1]{\operatorname{erf}}{(}{)}{}{#1}
\DeclarePairedDelimiterXPP\erfc[1]{\operatorname{erfc}}{(}{)}{}{#1}
\DeclarePairedDelimiterXPP\op[2]{\operatorname{#1}}{(}{)}{}{#2} % general operator

\newcommand{\T}{^{\intercal}}% transpose notation
\newcommand{\ud}{\,\mathrm{d}} % for integrals like \int f(x) \ud x
\newcommand{\bzero}{\mathbf{0}}
\newcommand{\bone}{\mathbf{1}}

\newcommand{\trace}[1]{{\Tr\left( #1 \right)}}

\newcommand{\KLD}[2]{{D({#1}\, \|\, {#2})}}

\providecommand\given{}
\newcommand\SetSymbol[2][]{%
\nonscript\, #1#2
\allowbreak
\nonscript\,
\mathopen{}}

\DeclarePairedDelimiterX\Set[2]\{\}{%
\renewcommand\given{\SetSymbol[\delimsize]{#1}}
#2
}
\DeclarePairedDelimiterX\Setc[1]\{\}{%
\renewcommand\given{\SetSymbol{:}}
#1
}

\NewDocumentCommand\set{s o m}{%
	\IfBooleanTF#1%
	{\IfValueTF{#2}{\Set*{#2}{#3}}{\Setc*{#3}}}%
	{\IfValueTF{#2}{\Set{#2}{#3}}{\Setc{#3}}}%
}

\NewDocumentCommand{\evalat}{s O{\big} m m}{%
  \IfBooleanTF{#1}
   {\left. #3 \right|_{#4}}
   {#3#2|_{#4}}%
}

\NewDocumentCommand \ifcond {m m} {%
	{#1} %
	\IfValueT{#2}{\, \middle|\, {#2}}%
}

\DeclareDocumentCommand \P {e{_} g >{\SplitArgument{ 1 }{ @| }}d() g } {%
	\mathbb{P}%
	\IfValueTF{#1}{_{#1}}
		{\IfValueT{#2}{_{#2}}}%
	\IfValueT{#3}{\left(\ifcond#3}%
	\IfValueT{#4}{\, \middle|\, {#4}}%
	\IfValueT{#3}{\right)}%
}

\DeclareDocumentCommand \E {e{_} g >{\SplitArgument{ 1 }{ @| }}o g } {%
	\mathbb{E}%
	\IfValueTF{#1}{_{#1}}
		{\IfValueT{#2}{_{#2}}}%
	\IfValueT{#3}{\left[\ifcond#3}%
	\IfValueT{#4}{\, \middle|\, {#4}}%
	\IfValueT{#3}{\right]}%
}

\newcommand{\N}[2]{{\calN\left({#1},\, {#2}\right)}}

\graphicspath{{./Figures/}} 
\newcommand{\includeCroppedPdf}[2][]{%
    \IfFileExists{./Figures/#2-crop.pdf}{}{%
        \immediate\write18{pdfcrop ./Figures/#2 ./Figures/#2-crop.pdf}}%
    \includegraphics[#1]{./Figures/#2-crop.pdf}}

\definecolor{gray90}{gray}{0.9}

\ifx\nohighlights\undefined

	\newcommand{\msout}[1]{\text{\color{green} \sout{\ensuremath{#1}}}}
	\newcommand{\del}[1]{{\color{green}\ifmmode \msout{#1}\else\sout{#1}\fi}}
\else

	\newcommand{\msout}[1]{#1}
	\newcommand{\del}[1]{#1}
\fi

\newcommand{\hhide}[1]{}

\newcommand{\txp}[2]{\texorpdfstring{#1}{#2}}

\ifx\diagnoselabel\undefined
	\relax
\else
	\makeatletter
	 \def\@testdef #1#2#3{%
		 \def\reserved@a{#3}\expandafter \ifx \csname #1@#2\endcsname
		\reserved@a  \else
	 \typeout{^^Jlabel #2 changed:^^J%
	 \meaning\reserved@a^^J%
	 \expandafter\meaning\csname #1@#2\endcsname^^J}%
	 \@tempswatrue \fi}
	\makeatother
\fi

%% file: MLPrivacy_v6.bbl
\begin{thebibliography}{10}
\providecommand{\url}[1]{#1}
\csname url@samestyle\endcsname
\providecommand{\newblock}{\relax}
\providecommand{\bibinfo}[2]{#2}
\providecommand{\BIBentrySTDinterwordspacing}{\spaceskip=0pt\relax}
\providecommand{\BIBentryALTinterwordstretchfactor}{4}
\providecommand{\BIBentryALTinterwordspacing}{\spaceskip=\fontdimen2\font plus
\BIBentryALTinterwordstretchfactor\fontdimen3\font minus
  \fontdimen4\font\relax}
\providecommand{\BIBforeignlanguage}[2]{{%
\expandafter\ifx\csname l@#1\endcsname\relax
\typeout{** WARNING: IEEEtran.bst: No hyphenation pattern has been}%
\typeout{** loaded for the language `#1'. Using the pattern for}%
\typeout{** the default language instead.}%
\else
\language=\csname l@#1\endcsname
\fi
#2}}
\providecommand{\BIBdecl}{\relax}
\BIBdecl

\bibitem{FirGolElo:J14}
M.~Fire, R.~Goldschmidt, and Y.~Elovici, ``Online social networks: {T}hreats
  and solutions,'' \emph{IEEE Communications Surveys Tutorials}, vol.~16,
  no.~4, pp. 2019--2036, May 2014.

\bibitem{AbaNinHer:J16}
J.~H. Abawajy, M.~I.~H. Ninggal, and T.~Herawan, ``Privacy preserving social
  network data publication,'' \emph{IEEE Communications Surveys Tutorials},
  vol.~18, no.~3, pp. 1974--1997, Mar. 2016.

\bibitem{WiKi:2018}
\BIBentryALTinterwordspacing
Wikipedia. 2018 {S}ing{H}ealth data breach. [Online]. Available:
  \url{https://en.wikipedia.org/wiki/2018_SingHealth_data_breach}
\BIBentrySTDinterwordspacing

\bibitem{CalFaw:C12}
F.~P. Calmon and N.~Fawaz, ``Privacy against statistical inference,'' in
  \emph{Proc. Allerton Conf. on Commun., Control and Computing}, Monticello,
  IL, Oct. 2012.

\bibitem{SunTay:J19b}
M.~Sun and W.~P. Tay, ``On the relationship between inference and data privacy
  in decentralized {I}o{T} networks,'' \emph{{IEEE} Trans. Inf. Forensics
  Security}, vol.~15, no.~1, pp. 852--866, Dec. 2020.

\bibitem{NyPap:J14}
J.~L. Ny and G.~J. Pappas, ``Differentially private filtering,'' \emph{{IEEE}
  Trans. Autom. Control}, vol.~59, no.~2, pp. 341--354, Feb. 2014.

\bibitem{SunTay:J19a}
M.~Sun and W.~P. Tay, ``Decentralized detection with robust information privacy
  protection,'' \emph{{IEEE} Trans. Inf. Forensics Security}, vol.~15, no.~1,
  pp. 85--99, Nov. 2020.

\bibitem{WanYinZha:J16}
W.~Wang, L.~Ying, and J.~Zhang, ``On the relation between identifiability,
  differential privacy, and mutual-information privacy,'' \emph{{IEEE} Trans.
  Inf. Theory}, vol.~62, no.~9, pp. 5018--5029, Sep. 2016.

\bibitem{Kung:J2017}
S.~Y. Kung, ``Compressive privacy: {F}rom information\/estimation theory to
  machine learning,'' \emph{{IEEE} Signal Process. Mag.}, vol.~34, no.~1, pp.
  94--112, Jan. 2017.

\bibitem{TseBoWu:J2020}
B.~O. Tseng and P.~Y. Wu, ``Compressive privacy generative adversarial
  network,'' \emph{{IEEE} Trans. Inf. Forensics Security}, vol.~15, pp.
  2499--2513, Jan. 2020.

\bibitem{SonWanTay:J19}
Y.~Song, C.~X. Wang, and W.~P. Tay, ``Compressive privacy for a linear
  dynamical system,'' \emph{{IEEE} Trans. Inf. Forensics Security}, vol.~15,
  no.~1, pp. 895--910, Dec. 2020.

\bibitem{SanRajPoo:J2013}
L.~{Sankar}, S.~R. {Rajagopalan}, and H.~V. {Poor}, ``Utility-{P}rivacy
  tradeoffs in databases: {A}n information-theoretic approach,'' \emph{{IEEE}
  Trans. Inf. Forensics Security}, vol.~8, no.~6, pp. 838--852, Jun. 2013.

\bibitem{LiaSanTan:J2017}
J.~C. Liao, L.~Sankar, V.~Y.~F. Tan, and F.~P. Calmon, ``Hypothesis testing
  under mutual information privacy constraints in the high privacy regime,''
  \emph{{IEEE} Trans. Inf. Forensics Security}, vol.~13, no.~4, pp. 1058--1071,
  Dec. 2017.

\bibitem{MakSalFaw:C14}
A.~Makhdoumi, S.~Salamatian, N.~Fawaz, and M.~M{\'e}dard, ``From the
  information bottleneck to the privacy funnel,'' in \emph{Proc. IEEE Inform.
  Theory Workshop}, Hobart, TAS, Australia, Nov. 2014.

\bibitem{WanCal:C2017}
H.~Wang and F.~P. Calmon, ``An estimation-theoretic view of privacy,'' in
  \emph{Proc. Allerton Conf. Commun., Control and Computing}, Monticello, IL,
  USA, Oct. 2017.

\bibitem{WanVoCal:J2019}
H.~Wang, L.~Vo, F.~P. Calmon, M.~M{\'e}dard, K.~R. Duffy, and M.~Varia,
  ``Privacy with estimation guarantees,'' \emph{{IEEE} Trans. Inf. Theory},
  vol.~65, no.~12, pp. 8025--8042, Aug. 2019.

\bibitem{CalMak:J2017}
F.~P. Calmon, A.~Makhdoumi, M.~Médard, M.~Varia, M.~Christiansen, and K.~R.
  Duffy, ``Principal inertia components and applications,'' \emph{{IEEE} Trans.
  Inf. Theory}, vol.~63, no.~8, pp. 5011--5038, Aug. 2017.

\bibitem{AsoAlaLin:C16}
S.~Asoodeh, F.~Alajaji, and T.~Linder, ``Privacy-aware {MMSE} estimation,'' in
  \emph{Proc. IEEE Int. Symp. on Inform. Theory}, Barcelona, Spain, Jul. 2016.

\bibitem{AsoDiaLin:J17}
S.~Asoodeh, M.~Diaz, F.~Alajaji, and T.~Linder, ``Estimation efficiency under
  privacy constraints,'' \emph{arXiv preprint arXiv:1707.02409}, Jul. 2017.

\bibitem{CalMak:C2015}
F.~P. Calmon, A.~Makhdoumi, and M.~Médard, ``Fundamental limits of perfect
  privacy,'' in \emph{Proc. IEEE Int. Symp. on Inform. Theory}, Hong Kong,
  China, Jun. 2015.

\bibitem{RasGun:J2017}
B.~Rassouli and D.~Gunduz, ``On perfect privacy and maximal correlation,''
  \emph{arXiv preprint arXiv:1712.08500}, 2017.

\bibitem{WanSonTay:J2020}
C.~X. Wang, Y.~Song, and W.~P. Tay, ``Arbitrarily strong utility-privacy
  tradeoff in multi-agent systems,'' \emph{{IEEE} Trans. Inf. Forensics
  Security}, pp. 1--1, 2020.

\bibitem{SonWanTay:C18}
Y.~Song, C.~X. Wang, and W.~P. Tay, ``Privacy-aware {K}alman filtering,'' in
  \emph{Proc. IEEE Int. Conf. Acoustics, Speech, and Signal Processing},
  Calgary, Canada, Apr. 2018.

\bibitem{WanSon:C2018}
C.~X. Wang, Y.~Song, and W.~P. Tay, ``Preserving parameter privacy in sensor
  networks,'' in \emph{Proc. IEEE Global Conf. on Signal and Information
  Processing}, Anaheim, USA, Nov. 2018.

\bibitem{ChaChaKun:C17}
T.~Chanyaswad, J.~M. Chang, and S.~Y. Kung, ``A compressive multi-kernel method
  for privacy preserving machine learning,'' in \emph{Proc. Int. Joint Conf.
  Neural Networks (IJCNN)}, Anchorage, Alaska, May 2017.

\bibitem{SunTayHe:J18}
M.~Sun, W.~P. Tay, and X.~He, ``Toward information privacy for the internet of
  things: {A} nonparametric learning approach,'' \emph{{IEEE} Trans. Signal
  Process.}, vol.~66, no.~7, pp. 1734--1747, Apr. 2018.

\bibitem{HeWee:J2019}
X.~He, W.~P. Tay, L.~Huang, M.~Sun, and Y.~Gong, ``Privacy-aware sensor network
  via multilayer nonlinear processing,'' \emph{IEEE Internet Things J.},
  vol.~6, pp. 10\,834--10\,845, Dec. 2019.

\bibitem{ChrKevYuj:J2015}
C.~Louizos, K.~Swersky, Y.~J. Li, M.~Welling, and R.~Zemel, ``The variational
  fair autoencoder,'' \emph{arXiv preprint arXiv:1511.00830v6}, 2017.

\bibitem{DanShuRob:J2018}
D.~Moyer, S.~Y. Gao, R.~Brekelmans, G.~V. Steeg, and A.~Galstyan, ``Invariant
  representations without adversarial training,'' \emph{arXiv preprint
  arXiv:1805.09458v3}, 2019.

\bibitem{RicYuKev:C2013}
\BIBentryALTinterwordspacing
R.~Zemel, Y.~Wu, K.~Swersky, T.~Pitassi, and C.~Dwork, ``Learning fair
  representations,'' in \emph{Proc. Int. Conf. on Machine Learning}, Atlanta,
  Georgia, USA, Jun. 2013. [Online]. Available:
  \url{http://proceedings.mlr.press/v28/zemel13.html}
\BIBentrySTDinterwordspacing

\bibitem{ChoPetXia:J2018}
C.~Huang, P.~Kairouz, X.~Chen, L.~Sankar, and R.~Rajagopal, ``Generative
  adversarial privacy,'' \emph{arXiv preprint arXiv:1807.05306v3}, 2019.

\bibitem{ChoPetXia:J2017}
------, ``Context-aware generative adversarial privacy,'' \emph{arXiv preprint
  arXiv:1710.09549v3}, 2017.

\bibitem{TriWanIsh:C19}
A.~Tripathy, Y.~Wang, and P.~Ishwar, ``Privacy-preserving adversarial
  networks,'' in \emph{Proc. Allerton Conf. on Commun., Control and Computing},
  Monticello, IL, USA, Sep. 2019.

\bibitem{ZhaSonQi:C2017}
Z.~F. Zhang, Y.~Song, and H.~R. Qi, ``Age progression/regression by conditional
  adversarial autoencoder,'' in \emph{Proc. IEEE Conference on Computer Vision
  and Pattern Recognition (CVPR)}, Hawaii, United States, Jul. 2017.

\bibitem{ErfHasJoh:J2004}
\BIBentryALTinterwordspacing
B.~Efron, T.~Hastie, I.~Johnstone, and R.~Tibshirani, ``Least angle
  regression,'' \emph{Annals of Statistics}, vol.~32, pp. 407--499, Apr. 2004.
  [Online]. Available: \url{https://projecteuclid.org/euclid.aos/1083178935}
\BIBentrySTDinterwordspacing

\bibitem{KinWel:J2013}
D.~P. Kingma and M.~Welling, ``Auto-encoding variational bayes,'' \emph{arXiv
  preprint arXiv:1312.6114}, 2013.

\bibitem{GreBor:J2006}
A.~Gretton, K.~M. Borgwardt, M.~Rasch, B.~Schoelkopf, and A.~J. Smola, ``A
  kernel method for the two-sample-problem,'' \emph{Advances in neural
  information processing systems}, pp. 513--520, 2006.

\bibitem{ZhuPar:J2018}
J.~Y. Zhu, T.~Park, P.~Lsola, and A.~A. Efros, ``Unpaired image-to-image
  translation using adversarial consistency loss,'' \emph{arXiv preprint
  arXiv:1703.10593}, 2018.

\bibitem{MurKolKri:C2018}
Z.~Murez, S.~Kolouri, D.~Kriegman, R.~Ramamoorthi, and K.~Kim, ``Image to image
  translation for domain adaptation,'' in \emph{arXiv preprint
  arXiv:1712.00479}, 2018.

\bibitem{Mul:J1997}
A.~M{\"u}ller, ``Integral probability metrics and their generating classes of
  functions,'' \emph{Advances in Applied Probability}, vol.~29, no.~2, pp.
  429--443, Jun. 1997.

\bibitem{SriFukLan:J2011}
\BIBentryALTinterwordspacing
B.~K. Sriperumbudur, K.~Fukumizu, and G.~R.~G. Lanckriet, ``Universality,
  characteristic kernels and {RKHS} embedding of measures,'' \emph{Journal of
  Machine Learning Research}, vol.~12, no.~70, pp. 2389--2410, Jul. 2011.
  [Online]. Available: \url{http://jmlr.org/papers/v12/sriperumbudur11a.html}
\BIBentrySTDinterwordspacing

\bibitem{ManAmb:J2015}
J.~H. Manton and P.~O. Amblard, ``A primer on reproducing kernel {H}ilbert
  spaces,'' \emph{arXiv preprint arXiv:1408.0952v2}, 2015.

\bibitem{GreBorRas:J2012}
\BIBentryALTinterwordspacing
A.~Gretton, K.~M. Borgwardt, M.~J. Rasch, B.~Sch{{\"o}}lkopf, and A.~Smola, ``A
  kernel two-sample test,'' \emph{Journal of Machine Learning Research},
  vol.~13, no.~25, pp. 723--773, Mar. 2012. [Online]. Available:
  \url{http://jmlr.org/papers/v13/gretton12a.html}
\BIBentrySTDinterwordspacing

\bibitem{LiChaChe:J2017}
C.~L. Li, W.~C. Chang, Y.~Cheng, Y.~M. Yang, and B.~Poczos, ``{MMD GAN}:
  {T}owards deeper understanding of moment matching network,'' in \emph{Proc.
  Int. Conf. on Neural Inform. Processing Systems}, Long Beach, CA, USA, Dec.
  2017.

\bibitem{LiSweZem:J2015}
Y.~Li, K.~Swersk, and R.~Zeme, ``Generative moment matching networks,''
  \emph{arXiv preprint arXiv:1502.02761}, 2015.

\bibitem{Hir:J1935}
H.~O. Hirschfeld, ``A connection between correlation and contingency,'' in
  \emph{Mathematical Proceedings of the Cambridge Philosophical Society},
  vol.~31.\hskip 1em plus 0.5em minus 0.4em\relax Cambridge University Press,
  1935, pp. 520--524.

\bibitem{Gel:J1941}
H.~Gebelein, ``Das statistische problem der korrelation als variations-und
  eigenwertproblem und sein zusammenhang mit der ausgleichsrechnung,''
  \emph{ZAMM-Journal of Applied Mathematics and Mechanics/Zeitschrift f{\"u}r
  Angewandte Mathematik und Mechanik}, vol.~21, no.~6, pp. 364--379, 1941.

\bibitem{Ren:J1959}
A.~R{\'e}nyi, ``On measures of dependence,'' \emph{Acta Math. Hung.}, vol.~10,
  pp. 441--451, Sep. 1959.

\bibitem{HuaZhe:J2014}
L.~Z. S.~L.~Huang, ``The linear information coupling problems,'' \emph{arXiv
  preprint arXiv:1406.2834}, 2014.

\bibitem{Con:B90}
J.~B. Conway, \emph{A Course in Functional Analysis}, 2nd~ed.\hskip 1em plus
  0.5em minus 0.4em\relax New York, NY: Springer-Verlag, 1990.

\bibitem{Sch:J1966}
\BIBentryALTinterwordspacing
P.~H. Sch{\"o}nemann, ``A generalized solution of the orthogonal procrustes
  problem,'' \emph{Psychometrika 31}, pp. 1--10, 1966. [Online]. Available:
  \url{https://doi.org/10.1007/BF02289451}
\BIBentrySTDinterwordspacing

\bibitem{BreFri:J1985}
L.~Breiman and J.~H. Friedman, ``Estimating optimal transformations for
  multiple regression and correlation,'' \emph{Journal of the American
  Statistical Association}, vol.~80, no. 391, pp. 614--619, Sep. 1985.

\bibitem{HarSzeSha:J2004}
D.~R. Hardoon, S.~Szedmak, and J.~Shawe-Taylor, ``Canonical correlation
  analysis: {A}n overview with application to learning methods,'' \emph{Neural
  computation}, vol.~16, no.~12, pp. 2639--2664, Dec. 2004.

\bibitem{BacJor:J2012}
F.~R. Bach and M.~I. Jordan, ``Kernel independent component analysis,''
  \emph{Journal of Machine Learning Research}, vol.~3, pp. 1--48, Jul. 2002.

\bibitem{SchSmoMul:J1998}
\BIBentryALTinterwordspacing
B.~Sch{\"o}lkopf, A.~Smola, and K.~R. M{\"u}ller, ``Nonlinear component
  analysis as a kernel eigenvalue problem,'' \emph{Neural Computation},
  vol.~10, no.~5, pp. 1299--1319, Jul. 1998. [Online]. Available:
  \url{https://doi.org/10.1162/089976698300017467}
\BIBentrySTDinterwordspacing

\bibitem{GreBouSmo:J2005}
A.~Gretton, O.~Bousquet, A.~Smola, and B.~Sch{\"o}lkopf, ``Measuring
  statistical dependence with {H}ilbert-{S}chmidt norms,'' in \emph{Algorithmic
  Learning Theory: ALT 2005}.\hskip 1em plus 0.5em minus 0.4em\relax Springer,
  2005, pp. 63--77.

\bibitem{FukBacJor:J2004}
\BIBentryALTinterwordspacing
K.~Fukumizu, F.~R. Bach, and M.~I. Jordan, ``Dimensionality reduction for
  supervised learning with reproducing kernel {H}ilbert spaces,'' \emph{Journal
  of Machine Learning Research}, vol.~5, pp. 73--99, Jan. 2004. [Online].
  Available: \url{http://www.jmlr.org/papers/v5/fukumizu04a.html}
\BIBentrySTDinterwordspacing

\bibitem{Bak:J1973}
\BIBentryALTinterwordspacing
C.~R. Baker, ``Joint measures and cross-covariance operators,''
  \emph{Transactions of the American Mathematical Society}, vol. 186, pp.
  273--289, Dec. 1973. [Online]. Available:
  \url{https://www.ams.org/tran/1973-186-00/S0002-9947-1973-0336795-3/}
\BIBentrySTDinterwordspacing

\bibitem{GreHerSmo:J2005}
\BIBentryALTinterwordspacing
A.~Gretton, R.~Herbrich, A.~Smola, O.~Bousquet, and B.~Sch{\"o}lkopf, ``Kernel
  methods for measuring independence,'' \emph{Journal of Machine Learning
  Research}, vol.~6, pp. 2075--2129, Dec. 2005. [Online]. Available:
  \url{http://www.jmlr.org/papers/v6/gretton05a.html}
\BIBentrySTDinterwordspacing

\bibitem{LonSheDar:C2015}
J.~Long, E.~Shelhamer, and T.~Darrell, ``Fully convolutional networks for
  semantic segmentation,'' in \emph{Proc. IEEE Conf. on Computer Vision and
  Pattern Recognition}, Boston, USA, Jun. 2015.

\bibitem{JohAlaLi:C16}
J.~Johnson, A.~Alahi, and F.~F. Li, ``Perceptual losses for real-time style
  transfer and super-resolution,'' in \emph{Proc. European Conf. on Computer
  Vision}, Amsterdam, The Netherlands, Oct. 2016.

\bibitem{GooAbaMir:C2014}
I.~J. Goodfellow, J.~P. Abadie, M.~Mirza, B.~Xu, D.~W. Farley, S.~Ozair,
  A.~Courville, and Y.~Bengio, ``Generative adversarial nets,'' in \emph{Proc.
  Int. Conf. on Neural Information Processing Systems}, Montreal, Canada, Dec.
  2014.

\bibitem{IsoZhu:J2018}
P.~Isola, J.~Y. Zhu, T.~Zhou, and A.~A. Efros, ``Image-to-image translation
  with conditional adversarial networks,'' \emph{arXiv preprint
  arXiv:1611.07004v3}, 2018.

\bibitem{KaiXiaSha:J2015}
K.~He, X.~Zhang, S.~Ren, and J.~Sun, ``Deep residual learning for image
  recognition,'' \emph{arXiv preprint arXiv:1512.03385v1}, 2015.

\bibitem{ZemWu:C2013}
\BIBentryALTinterwordspacing
R.~Zemel, Y.~Wu, K.~Swersky, T.~Pitassi, and C.~Dwork,
  ``\BIBforeignlanguage{en}{Learning fair representations},'' in
  \emph{\BIBforeignlanguage{en}{Proc. Int. Conf. on Machine Learning}}.\hskip
  1em plus 0.5em minus 0.4em\relax PMLR, May 2013, pp. 325--333. [Online].
  Available: \url{http://proceedings.mlr.press/v28/zemel13.html}
\BIBentrySTDinterwordspacing

\bibitem{DuaGra:2019}
\BIBentryALTinterwordspacing
D.~Dua and C.~Graff. (2019) {UCI} machine learning repository. University of
  California, Irvine, School of Information and Computer Sciences. [Online].
  Available: \url{http://archive.ics.uci.edu/ml}
\BIBentrySTDinterwordspacing

\bibitem{Kung:J2017b}
\BIBentryALTinterwordspacing
S.~Y. Kung, T.~Chanyaswad, J.~M. Chang, and P.~Wu, ``Collaborative {PCA}/{DCA}
  learning methods for compressive privacy,'' \emph{ACM Trans. on Embed.
  Comput. Syst.}, vol.~16, no.~3, Jul. 2017. [Online]. Available:
  \url{https://doi.org/10.1145/2996460}
\BIBentrySTDinterwordspacing

\end{thebibliography}
